\documentclass[12pt,preprint]{aastex}
\usepackage{amsmath,amssymb,rotating} 
\newcommand{\mapiii}{MAPPINGS  \textsc{iii}}
\newcommand{\lam}{\ensuremath{\lambda}}
\newcommand{\mum}{\ensuremath{\mu\mbox{m}}}
\newcommand{\pccm}{\ensuremath{\,\mbox{cm}^{-3}}}
\newcommand{\kms}{\ensuremath{\,\mbox{km}\,\mbox{s}^{-1}}}
\newcommand{\flux}{\ensuremath{\,\mbox{erg}\,\mbox{cm}^{-2}\mbox{s}^{-1}}}
\catcode`\@=11
\def\gapprox{\mathrel{\mathpalette\@versim>}}
\def\lapprox{\mathrel{\mathpalette\@versim<}}
\def\@versim#1#2{\lower2.45pt\vbox{\baselineskip0pt\lineskip0.9pt
     \ialign{$\m@th#1\hfil##\hfil$\crcr#2\crcr\sim\crcr}}}
\catcode`\@=12
\newcommand{\hi}{H\,{\sc i}}
\newcommand{\hii}{H\,{\sc ii}}
\newcommand{\hei}{He\,{\sc i}}
\newcommand{\heii}{He\,{\sc ii}}
\newcommand{\heiii}{He\,{\sc iii}}
\newcommand{\ci}{C\,{\sc i}}
\newcommand{\cii}{C\,{\sc ii}}
\newcommand{\ciii}{C\,{\sc iii}}
\newcommand{\civ}{C\,{\sc iv}}
\newcommand{\cv}{C\,{\sc v}}
\newcommand{\cvi}{C\,{\sc vi}}
\newcommand{\cvii}{C\,{\sc vii}}
\renewcommand{\ni}{N\,{\sc i}}
\newcommand{\nii}{N\,{\sc ii}}
\newcommand{\niii}{N\,{\sc iii}}
\newcommand{\niv}{N\,{\sc iv}}
\newcommand{\nv}{N\,{\sc v}}
\newcommand{\oi}{O\,{\sc i}}
\newcommand{\oii}{O\,{\sc ii}}
\newcommand{\oiii}{O\,{\sc iii}}
\newcommand{\oiv}{O\,{\sc iv}}
\newcommand{\ov}{O\,{\sc v}}
\newcommand{\ovi}{O\,{\sc vi}}
\newcommand{\nei}{Ne\,{\sc i}}
\newcommand{\neii}{Ne\,{\sc ii}}
\newcommand{\neiii}{Ne\,{\sc iii}}
\newcommand{\neiv}{Ne\,{\sc iv}}
\newcommand{\nev}{Ne\,{\sc v}}
\newcommand{\nevi}{Ne\,{\sc vi}}
\newcommand{\mgii}{Mg\,{\sc ii}}
\newcommand{\siliv}{Si\,{\sc iv}}
\newcommand{\si}{S\,{\sc i}}
\newcommand{\sii}{S\,{\sc ii}}
\newcommand{\siii}{S\,{\sc iii}}
\newcommand{\siv}{S\,{\sc iv}}
\newcommand{\sv}{S\,{\sc v}}
\newcommand{\ari}{Ar\,{\sc i}}
\newcommand{\arii}{Ar\,{\sc ii}}
\newcommand{\ariii}{Ar\,{\sc iii}}
\newcommand{\ariv}{Ar\,{\sc iv}}
\newcommand{\arv}{Ar\,{\sc v}}
\newcommand{\arvi}{Ar\,{\sc vi}}
\newcommand{\fei}{Fe\,{\sc i}}
\newcommand{\feii}{Fe\,{\sc ii}}
\newcommand{\feiii}{Fe\,{\sc iii}}
\newcommand{\fevii}{Fe\,{\sc vii}}
\begin{document}
\title{Dusty, Radiation Pressure Dominated Photoionization. \\
I. Model Description, Structure And Grids}

\author{Brent A. Groves, Michael A. Dopita, \& Ralph S. Sutherland}

\affil{Research School of Astronomy \& Astrophysics,
Institute of Advanced Studies, The Australian National University,
Cotter Road, Weston Creek, ACT 2611
Australia}
\email{bgroves@mso.anu.edu.au, Michael.Dopita@anu.edu.au,
ralph@mso.anu.edu.au}

\begin{abstract}
We present the implementation
of dusty, radiation pressure dominated photoionization models
applicable to the Narrow Line Regions (NLRs) of Active Galactic
Nuclei, using the \mapiii\ code. We give a grid of the predicted
intensities of 
the most commonly used diagnostic spectral lines in the UV,
optical and IR, covering a  wide range of density, metallicity, the
power-law index characterizing the photoionizing source
and photoionization parameter, for use in the diagnosis
of NLRs. We examine the temperature, density and ionization structure
of these models, investigating the effect of variation of these
parameters in order to gain a better understanding of NLR clouds
themselves. 

\end{abstract}



\section{Introduction}

Emission line spectroscopy combined with photoionization or shock
modeling codes has proved to be an important tool in determining the
physical properties and chemical abundances of astrophysical
nebulae. The Narrow Line Regions (NLRs) within active galaxies
represent one of the more energetic types of ionized nebulae, and are
characterized by emission lines across a wide range of ionization
states. In addition, these regions are often spatially resolved in
both nearby and high-redshift galaxies as extended narrow line regions
(ENLRs). Resolved NLRs are excellent laboratories to study the
detailed physics of AGN nebulae, and allow strong tests of the various
ionization models which have been developed to explain their
properties.

Photoionization models invoke a nuclear source of high energy photons
to provide the dominant mechanism of ionization within a nebulae.
These models can be characterized by the dimensionless ionization
parameter $U = S_*/n_\mathrm{H}c$, where $S_*$ is the number of
ionizing photons passing through a unit area control surface per unit
time, $n_\mathrm{H}$ is the hydrogen density, and $c$ is the speed of
light. $U$ is therefore the ratio of the photon density to the atomic
density.  For active galaxies, the ``unified model'' holds that an
optically-thick torus confines the UV ionizing photons originating
from the central source (the active galactic nucleus or AGN) to a
cone-shaped region along the polar axis, and it is these photons that
excite the NLR.  The exact production mechanism of these photons
remains unclear, but to match the models to the observations the
ionizing spectrum is required to be a smooth, featureless power-law or
broken power-law
\citep{Koski78,Stasinska84,Ost89,Laor97,Alex00,Prieto00,Sazanov04}
with $U \sim 0.01$. This spectrum and the low densities (n$_\mathrm{H}
\sim 10^2 - 10^4$ \pccm) result in forbidden line emission over a wide
range of ionization states. Penetrating soft X-ray photons also ensure
the existence of an extensive partially ionized region in which the
lines of low-ionization species such as [\oi] are strong.  Both of
these properties are required by the observational constraints from
diagnostic emission line ratios involving low and intermediate
ionization states \citep{Veilleux87,Osterbrock92}.

However, simple constant-density photoionization models have
difficulty in explaining the existence of the strong coronal line
intensities relative to to H$\beta $, which are found in a number of
Seyfert galaxy NLR regions \citep{oliv94,moor96,KC2000}. The existence
of coronal gas requires that both the ionization parameter and the
radiation pressure are very large \citep{Ami96}. This is in
contradiction to the constraints on the ionization parameter provided
by the lower ionization lines which require $U \sim 0.01$
\citep{Veilleux87,Osterbrock92,VV00}.

This problem can be alleviated by abandoning the single density
(isochoric) photoionization models, and invoking multi-component
models of varying degrees of complexity, such as the $A_\mathrm{M/I}$
models of \citet{Ami96,Ami97} or the integrated emission models of
\citet{KomSch97} and \citet{Ferg97}. An unsatisfactory aspect of such
models is that the geometry of the absorbers with respect to the
ionization sources is a free parameter, so that the predictive nature
of the photoionization models is, to an extent, lost. HST/STIS
observations have been able to resolve several Seyfert galaxies down
to a few 10's of parsecs \citep[eg][]{Kraemer00,Bradley03}, and thus
place some constraints upon the geometry of these models, but not
enough to completely substantiate these models.

None of these photoionization models address the issue of the gas
dynamics of the photoionized region, which often display evidence of
non-gravitational motions \citep{Whittle96}. Examples of this can be
found amongst the more radio-luminous galaxies including Mrk78
\citep{Pedlar89}, NGC2992 \citep{Allen99} or Mrk 1066
\citep{Bower95}. Detailed HST/STIS observations of NLR also show strong
evidence of non-gravitational motions
\citep[eg][]{Crenshaw00,Kaiser00,Cecil02}.  These observations provide
the motivation to consider excitation by fast radiative shocks
\citep{DoSu95,DoSu96}. These are successful in reproducing many of the
observed diagnostic line ratios. In practice, these models are also
dominated by photoionization effects because shocks with velocities
$V_S \ge 200$ \kms\ generate a strong internal photoionizing field as
the shocked gas cools, which can ionize both the material ahead of the
shock as well as the cooler gas behind the shock. However, for Seyfert
galaxies at least, recent X-ray data \citep{Brinkman02, Ogle03}
provides powerful evidence for the existence of central photoionizing
fields with very high ionization parameters, and so makes such models
less viable, insofar as it is shocks which are the strongest source of
ionizing photons. Shocks may be still necessary to provide the
observed dynamical structure of many sources.

In an attempt to self-consistently address both the problems of the
gas dynamics and of the wide range in observed excitation conditions
that exist within the NLR, a new paradigm of dusty, radiation pressure
dominated photoionization was developed \citep{DG02} (DG02). This
model is based upon the standard photoionization paradigm, but it also
includes the physics of dust, known to exist in these regions, and
radiation pressure. These provide the physical basis needed to explain
the observed emission line ratios from NLRs, especially the
combination of high- and low- ionization lines within one physical
structure, and can also allow for the possibility of fast,
radiatively-driven outflows in Seyfert galaxies.

Here we continue the work of DG02 and explain how this model is
implemented in the photoionization \& shock code \mapiii. The majority
of the code has been discussed previously
\citep{Dopita82,Sutherland93}, and a general overview of the
photoionization calculations as implemented in the current code is
given in \citet{map3}. Here, we discuss only our implementation of
dust and radiation pressure within the code.

We also extend the previous work (DG02) by presenting the results of a
grid of both the standard photoionization models (with no dust or
radiation pressure, one set allowing for depletion of heavy elements
from the gas phase, and one set non-depleted) and our dusty, radiation
pressure dominated models.  These models cover a wide range in
metallicity, density, photoionization power-law index and ionization
parameter $U$.

The purpose of this paper is twofold. The first purpose is to present
the grid of models and the resulting line intensities, since complete
FUV-FIR diagnostics have not been previously presented, even for the
standard photoionization models.  The increasing availability of
emission line data, and, in particular, the launch of Spitzer makes such
a set of models timely as a diagnostic tool for observers.  Secondly,
we aim to provide insight into the physical structure of the
models. In association with the emission lines this assists in
understanding of how variation in the parameters results in the
observed ratios.  The full discussion of the emission line ratios, and
a detailed comparison of the models with available observations is
given in the following paper in this series \citep{Groves03}.

\section{Dust Models}\label{sec:dustmodels}

Dust is almost ubiquitous in the interstellar medium (ISM). It makes
its presence felt directly through its absorption, scattering, and
polarization of incident light. Warm dust can be seen directly through
its IR re-emission and its electric and magnetic dipole emission at
sub-mm wavelengths \citep{DraineLaz98}. Its existence can also be
indirectly inferred through the measurement of atomic depletion of
refractory species from the gas phase along various sight-lines
through the ISM.  These observations provide a wide number of
constraints on size distributions, chemical compositions and physical
state of the dust particles, yet even with these constraints, there
remains considerable uncertainty as to its form, physical structure
and chemical composition.

There are two main problems with distinguishing the general properties
of dust. First, dust properties are seen to vary between different
sight-lines and within different phases of the interstellar
medium. Such variation is a natural result of the combination of grain
growth (within dense dust clouds, or in the stellar outflows and
ejecta) and a wide variety of grain destruction or shattering
mechanisms.  Second, several of the observable properties of dust
appear to be degenerate in that they can be equally well reproduced by
different theoretical grain size distributions and compositions.

In the light of this complexity, we have elected to use in our
photoionization modeling the simplest grain composition and grain size
distribution model that is consistent with the known properties of the
dust in the solar neighborhood. This model is almost certainly
 not applicable
in detail to the environment of AGN, but it at least allows us to
evaluate the (often fairly gross) effects that dust will have on the
emission line spectrum of photoionized NLR. Additionally, the results
we have obtained and presented below should be easily reproducible by
other modeling codes.

It might be naively expected that the energetic environment and
powerful radiation fields known to exist in the vicinity of AGN would
be inimical to the survival of dust. However, dust is known to exist
within AGN, and spectral constraints on both its form and composition
have been derived \citep{Laor93}.

Most investigations of the dust content have concentrated upon the
dusty tori, however there is also evidence that dust exists in the NLR
of AGN \citep{Tomono01,Radomski03}. As discussed in \citet{DG02}, dust
survives in the dense molecular clouds and is not exposed the
radiation field from the AGN until it passes through the ionization
front surrounding these photoablating clouds. Despite the energetic
EUV field, refractory grain species are not primarily destroyed by
sublimation, but rather, by grain shattering followed by sublimation
of the smallest grains. The larger dust grains should be able to
survive their crossing of the high emissivity regions close to the
surface of NLR clouds where most of the UV, optical and IR lines are
produced.  However, the coronal species are produced further out in
the ionized flow, and dust is likely to be destroyed in these
regions. We therefore caution the reader against the use of coronal
line intensities predicted in the models presented in this paper, in
which grain destruction mechanisms are not taken into account.

\subsection{Dust Composition}

Through observations of the IR emission bands or UV absorption
features associated with dust, several particular dust species have
been identified in the ISM.  These include diamond \citep{VanK02},
crystalline silicates (such as enstatite MgSiO$_3$)\citep{Honda03},
water ice\citep{Alex03}, carbides (SiC, MgC, TiC)\citep{Hony03},
sulfides(FeS, MgS, SiS$_2$) \citep{Hony02} and carbonates (CaCO$_3$)
\citep{Onaka03}. Another important dust component which is believed to
play a major part in the physics of the neutral medium are polycyclic
aromatic hydrocarbons (PAHs). Their effect is felt through processes
such as photoelectric heating, and they are believed to be the source
of the ``unidentified bands'' (UIBs) in the infrared.

However, the existence and relative abundance of all these components
in the photoionized plasma around AGN is entirely unknown.  Therefore,
for simplicity, we use only two types of dust, amorphous
``astronomical'' silicates (such as forsterite or fayalite) and
amorphous carbon grains. Both are observed in the interstellar medium
through absorption and, together they can reproduce most of the
spectral properties of dust, including both the observed IR continuum
emission and the UV extinction \citep{Draine03}. It is these two
components which are the basis for most previous dust modeling, and
for which a large set of optical data exist.

The ``astronomical" amorphous silicate grains are generally described
as an olivine ((Mg,Fe)$_2$SiO$_4$) ranging somewhere between fayalite
(Fe$_2$SiO$_4$) and forsterite (Mg$_2$SiO$_4$), although some form of
enstatite has also been used. All of these have been observed in the
crystalline form, but the amorphous form allows for a random
structure, with possible vacuum spaces within the grain. This can
provide a broader emission without production of the crystalline
features which in any case are not observed in the ISM at
large\citep{Jager03}.

The amorphous carbon component is believed to range somewhere between
the less saturated aromatic forms characteristic of PAHs and a
saturated hydrocarbon form. Dust of such a composition is absolutely
necessary if we are to successfully reproduce the both the extinction
and emission data. This component is believed to be the source of the
2200\AA\ absorption band in the UV.

Although we have assumed only these two types of dust grain, we
deplete all elements which have gas-phase condensation temperatures
above a few hundred degrees.  The choice of the depletion factors is
based upon observations of the diffuse ISM \citep[eg.][]{SavageS96}.

\subsection{Dust Abundance}

Dust formation depletes the heavy elements from the surrounding gas.
Those that have a higher condensation temperature are observed to be
more depleted, while the noble gases such as He, Ne and Ar remain in
the gaseous phase. Although there is clear evidence that the depletion
of metals from the gas onto dust is strongly dependent on on the local
conditions \citep{SavageS96}, for simplicity we assume that the
fractional depletion of dust from the gas is the same as found within
the local interstellar medium. We also assume that this fractional
depletion does not vary with changes in the metallicity of the gas, so
that the gas-to-dust mass ratio is simply proportional to metallicity.

Such assumptions are a gross oversimplification, and cannot be
justified. We adopt them simply because, faced with our minimal
knowledge of dust depletion in active galaxies, they represent the
simplest and best understood grain model we can test, and they
minimize the abundance dependent effects of dust changing with
metallicity.

The adopted depletion factors are listed in table
\ref{tab:depl}. These values are those used by \citet{DopitaHII} for
starburst and active galaxy photoionization modeling and are similar
to those found by \citet{Jenkins87} and \citet{SavageS96} in the local
ISM using the UV absorption lines to probe various local lines of
sight.


The heavy element abundances and their depletion factors then
determines the total gas-to-dust ratio. The total mass of carbon
depleted from the gas phase determines the amount of carbonaceous
dust, while the remainder of the heavy elements locked up in dust
(mainly Si and O, but also including Fe, Ca and Mg) is assumed to have
gone into forming the silicates.  Although again, this is an
oversimplification, nonetheless we find that our model gives both a
reasonable ratio of carbon to silicates and an extinction generally
similar to that found within our galaxy. The total amount of dust is
then distributed across a range of sizes determined by the grain size
distribution.

\subsection{Grain Size Distribution}

The grain size distribution and the grain composition are the most
important factors determining the wavelength-dependence of the
absorption and scattering processes. The grain size distribution in
the ISM results from the balance between formation and destruction
processes. In general, it can be reasonably well-represented by a
power law over a wide range of sizes, $a$, \citet{MRN77} (MRN);
\begin{equation}
dN(a)/da = k a^{-\alpha} \quad a_\mathrm{min} \le a \le a_\mathrm{max}.
\end{equation}
More complex distributions have been suggested on physical grounds
\citep{Weingartner01a}, but even these can be approximated by
power-laws over a wide range of radii. Grain shattering has been shown
to lead naturally to the formation of a power-law size distribution of
grains with $\alpha \sim 3.3$ \citep{Jones96}, observationally
indistinguishable from the MRN value, $\alpha \sim 3.5$.

We have adopted MRN grain size distribution in our models with $\alpha
\sim 3.5$ and the factor $k$ being determined by the gas-to-dust mass
ratio. For silicate grains we take $a_\mathrm{min} = 100$ \AA\ and
$a_\mathrm{max} = 2500$ \AA, while for carbonaceous grains we have
$a_\mathrm{min} = 50$ \AA\ and $a_\mathrm{max} = 2500$ \AA. These
lower limits are considerably larger than might be used in the local
ISM, since we expect that in AGN environments small grains are much
more likely to be destroyed.  This absence of small grains reduces
both the total absorption and the photoelectric heating by grains in
the models.  The assumption of a large minimum grain size is supported
by the relative absence of the 2200 \AA\ dust feature in AGN spectra
\citep{Pitman00}. This feature is believed to be due to graphite dust,
and is dominated by the smallest grain sizes. So an absence of this
feature indicates that the smallest of grains are insignificant in
quantity.

Within the \mapiii\ code, we divide the dust grains into 80 bins
spaced logarithmically between 0.001--10 \mum. The number in each bin
is then determined by the distribution. The absorption, scattering and
photoelectric heating within the code is then calculated for each size
bin.

\subsection{Extinction and Scattering by Dust}

For the optical properties of the two compositions of dust we use the
data created by Draine and co-authors using calculated dielectric
functions and a combination of Mie and Rayleigh-Gans theory.

To represent the carbon grains we have used the graphite optical data
\citep{DraineLee84,Laor93} and for the silicate grains the ``smoothed
astronomical silicate'' data \citep{DraineLee84,Laor93,
Weingartner01a}. These data\footnote{The optical data can be
downloaded at
http://www.astro.princeton.edu/$\sim$draine/dust/dust.diel.html }
include the scattering and absorption cross-sections for grains
ranging from 0.001--10 \mum.  A full description of the data and its
construction can be found in the cited papers.

In \mapiii\ the data is read in as a table and applied to each
relevant grain size bin for both the graphitic and silicaceous grains,
allowing the absorption and scattering cross-section to be calculated
for each bin. The final total extinction and scattering cross-section
of dust within the models is shown in figure \ref{fig:ext}.

\placefigure{Fig.~\ref{fig:ext}}

\subsection{Photoelectric Heating}

Photoelectric heating of the gas by dust has been shown to be an
important factor in influencing the temperature structure of both
\hii\ regions \citep{Maciel82} and planetary nebulae
\citep{Borkowski91}.  This process was first computed quantitatively
by \citet{Draine78}, and this work has been followed by more accurate
calculations which explore both the physics of the process
\citep{Weingartner01b} and explore the importance of the photoelectric
(PE) effect in nebular modeling \citep{DopitaPE}.

The charge of dust grains is determined by the balance between the PE
effect and the collisional `sticking' of charged particles to the
grain. In highly ionized nebulae with strong radiation fields, the PE
effect dominates and grains are positively charged, but when the
radiation field is weak electron sticking dominates and the grains are
negatively charged.

The implementation of the photoelectric effect on dust in \mapiii\ has
already been discussed previously in \citet{DopitaPE}.  This
implementation has since been altered so that the heating is
calculated for the individual grain size bins rather than for the dust
as a whole. This allows different grain sizes to have different
charges.

The photoelectric heating turns out to be important in our NLR models
as it strongly heats and excites the high-ionization gas, and helps
address the temperature problem of these regions.  Charged grains are
locked to the ionized plasma (drift velocities are only a few
cm~sec$^{-1}$), so the charge serves to couple the radiation pressure
(acting on dust) directly to the gas \citep{DG02}.

\subsection{Radiation Pressure}

The importance of radiation pressure acting on dust in high ionization
parameter photoionized regions was demonstrated in DG02. In an ionized
plasma the radiation pressure forces can be separated into two
components: the force exerted by the photoionization process and the
component due to the absorption by dust,
\begin{eqnarray}
F_{{\rm rad}}(x)=&&\sum_{m=1}{\sum_{i=0}{n_{i}(X_{m}^{+i})\int_{\nu
_{m}^{i}}^{\infty }{\frac{{\varphi (\nu )}}{{c}}}a_{\nu }(X_{m}^{+i})~d\nu
}}
\nonumber \\
&+&n_{H}\int_{0}^{\infty } \left[\kappa_{{\rm abs}}\left( \nu
\right)+\kappa_{{\rm sca}}\left ( \nu \right)
\left(1-g\left(\nu\right)\right)\right]{\frac{{\varphi (\nu )}}{{c}}}d\nu~,
\label{eqn:radforce}
\end{eqnarray}
where $\varphi (\nu )$ is the local radiation flux at frequency $\nu$,
$n_{i}(X_{m}^{+i})$ the number density of ion $+i$ of atomic species
$m$, $a_{\nu }(X_{m}^{+i})$ the corresponding photoionization cross
section with threshold $\nu _{m}^{i}$. In the second part, the dust
term, $\kappa_{{\rm abs}}\left( \nu \right) $ and $\kappa_{{\rm
sca}}\left( \nu \right) $ are the dust absorption cross section and
scattering cross section respectively, both normalized to the hydrogen
density, and $g\left(\nu\right) = \left\langle \cos \theta
_{\nu}\right\rangle $ is the mean scattering angle of the dust grains
at this frequency. Note that the major component of the radiation
pressure on dust comes from the far-UV region of the spectrum, where
the dust opacity is greatest and the opacity due to hydrogen is
effectively zero. The radiation force acting directly onto the gas can
be considered solely due to photoionization, as the bound-bound
transitions contribute a relatively negligible force due to the
stationary nature of the NLR cloud gas along the stagnation point.

Equation \ref{eqn:radforce} contains the implicit assumption that the
gas and the dust are closely coupled by Coulomb forces and hence any
force acting upon the dust can be considered as imparted to the system
as a whole.

In the case of an ionization front facing the source of ionizing
photons, the ionized plasma can be approximated as being in
hydrostatic equilibrium (which is assumed in our isobaric models), and
in this case the gas pressure gradient must match the local radiative
volume force,
\begin{equation}
\frac{dP_\mathrm{gas}}{dx} = F_{{\rm rad}}(x).
\end{equation}
This gives the gas pressure structure of the nebulae, from which the
density structure can be determined, assuming equilibrium conditions.
For ionization fronts which are oblique to the ionizing source, the
radiation pressure provides a shear force which serves to accelerate
the flow away from the source of UV photons. In this case, a
hydrodynamic photoionization model would be required to describe the
flow. This is beyond the scope of this paper.

\subsection{Grain Destruction}

The assumption of close coupling between gas and dust breaks down at a single point in the
flow from the ionization front where the collisional charging and the photoelectric
charging rates of a grain balance, and the grain consequently has zero charge.
As these processes are size dependent, this point occurs at a
different part of the flow for each grain size.  At this point, the
full radiation pressure force acts on the grain, and it is rapidly
accelerated towards its terminal velocity with respect to the gas and other grain sizes. The terminal velocity is 
determined by the kinetic friction forces exerted on it by the
gas. This phenomenon is almost certainly unimportant in affecting the
ionized gas flow as a whole, but it could be important in providing a
means of shattering grains by high-velocity grain-grain collisions.

As this process occurs at different points for each grain size, with the largest grains becoming neutral first in the flow, what can occur is that the large grains become neutral, dissociate from the gas, and are shattered to small grains which become charged again. These then become neutral further down in the photoablated flow from the NLR cloud and the process is repeated. This will reoccur until the grains are small enough to be destroyed through stochastic heating. Thus through this process we have a mechanism to destroy dust grains and return the metals in dust to the gas in the coronal region around NLR clouds. Modelling this destruction however, is beyond the scope of this work.

\section{Chemical Abundances}\label{sec:abund}

The major parameters which influence the spectrum of a photoionized
plasma are the ionization parameter and the chemical abundances.  To
investigate the influence of abundance, we use five abundance sets; $4
\times \mathrm{Z}_\odot$, $2 \times \mathrm{Z}_\odot$, $
\mathrm{Z}_\odot$, $0.5 \times \mathrm{Z}_\odot$, and $0.25 \times
\mathrm{Z}_\odot$, where $\mathrm{Z}_\odot$ represents solar
metallicity.

\subsection{The Solar Abundance Set}\label{sec:solarabund}
The last few years have seen a more accurate determination of the
solar abundances of several chemical species, including carbon,
nitrogen and oxygen. These papers by Asplund and his collaborators
\citep{AsplundNord00,Asplund00,AllendeLamb01,AllendeLamb02,Asplund03}
use 3D hydrodynamical modeling of the solar atmosphere along with
detailed atomic and molecular data to reanalyze high S/N Solar spectra
and so obtain revised solar abundances for C, N, O, Si and Fe. These
abundances are generally lower than the previous estimates by
\citet{Anders89} and are are quite similar to those found in the local
ISM and surrounding B stars \citep{Russell89,Russell90,Russell92}.

We have adopted the C, N, O, Si and Fe abundances from Asplund et~al.,
and the abundances of the remaining elements in our solar abundance
set, excluding the noble gases, are obtained from the meteoritic
abundances in the standard solar abundance set of
\citet{Grevesse98}. Ne and Ar are obtained from the photospheric
measurements of \citet{Grevesse98}.

Due to its absence from the solar photospheric spectrum, the He
represents a special case. We have used the primordial measurements of
\citet{Pagel92} along with their linear fits of $Y$ with [O/H] to
estimate the solar Helium abundance.

The final Solar abundance set used within our models closely resembles
that observed in the local ISM \citep{Russell89,Russell90,Russell92}
and is given in table \ref{tab:solar}.


\subsection{Scaling the Solar Abundance Set}
For primary nucleosynthetic elements, their abundance with respect to
hydrogen scales with overall metallicity. The two exceptions to this
relationship are He and N.

For Helium, the chemical yield from stars simply adds to the
primordial abundance estimated by \citet{Pagel92} to be He/H = 0.0737
(Y$_P$ = 0.228).  This gives:
\begin{equation}\label{eqn:He/H}
\mathrm{He}/{\mathrm{H}} = 0.0737 +
0.0293\mathrm{Z}/{\mathrm{Z}_\odot}.
\end{equation}

Nitrogen is well known to possess both a primary and secondary
nucleosynthetic component. For low metallicity galaxies (log (O/H)
$\le -4.0$) the N/O ratio is approximately constant, as expected for a
``primary'' element whose production is independent of
metallicity. However, for high metallicity galaxies (log (O/H) $\ge
-3.5$) the N/O ratio is found to rise steeply with metallicity
suggesting that nitrogen becomes a ``secondary'' element whose
production is proportional to the abundance of other heavy elements,
mainly C and O.

To take account of this effect, we have created a linear combination
of the primary and secondary components of Nitrogen, fitting the
resulting this curve to several abundance data sets, with the only
condition being that it must pass through the Solar abundance values.

Three data sets were used, with two of these coming from
\citet{Mouhcine02}; the \hii\ Galaxy sample and the starburst nuclear
galaxy sample (SBNG), both of which are compilations of several
surveys. The third data set comes from \citet{Kennicutt03} which are
measurements of \hii\ regions within the giant spiral galaxy M101.

\placefigure{Fig.~\ref{N/Ofig}}

The analytic function obtained for the nitrogen abundance is
\begin{equation}\label{eqn:N/H}
\left[\mathrm{N}/\mathrm{H}\right] =\left[\mathrm{O}/\mathrm{H}\right]
\left(10^{-1.6} + 10^{\left(2.33 +
\log_{10}\left[\mathrm{O}/\mathrm{H}\right]\right)}\right)
\end{equation}

The fit of this curve with data can be seen in figure
\ref{N/Ofig}. The fit is generally good but as the curve must pass
through the solar values, it tends to give a higher N/O ratio than the
observations at large metallicities.

\section{Photoionization Models}\label{sec:photomodels}

In \citet{DG02} we demonstrated the important effects of both
radiation pressure and dust on the output line ratios in
photoionization models. In this and a following paper \citep[Paper
2]{Groves03} we will systematically explore the parameter space of the
ionization parameter, abundance, slope of power-law photoionizing
spectrum, and density for both dusty, radiation pressure dominated
photoionization models and for the standard (dust and
radiation-pressure free) photoionization models. The chosen physical
parameters are common to both types of model and are chosen to cover
the typical values inferred for the NLR of Active Galaxies.

The photoionization models are all one dimensional, equilibrium models
and have a plane parallel geometry, representing the NLR clouds far
from the AGN. Both sets of models are truncated at the point at which
the fraction of \hii\ drops below 1\%.

\subsection{Abundance Variation}

As stated in the previous section (\S \ref{sec:abund}) we have run
five abundance sets with metallicities of $~0.25 \mathrm{Z}_\odot,~
0.5 \mathrm{Z}_\odot,~ 1 \mathrm{Z}_\odot,~ 2 \mathrm{Z}_\odot$ and $4
\mathrm{Z}_\odot$ with relative abundances as discussed in that
section.

The gas abundances in both the dusty and dust-free models are depleted
by dust (by the amounts given in Table \ref{tab:depl}). In the dusty,
radiation pressure models the depleted heavy elements reside in the
dust, but in the standard photoionization models the depleted heavy
elements are effectively lost. The reason for this is that the cooling
and, to some extent, the ionization are determined by the gas phase
abundance of heavy elements. It is therefore important for comparison
purposes that the gas phase abundances in both sets of models is the
same. To demonstrate this effect we have run a third set of standard,
dust-free photoionization models which are not depleted. This third
set reproduces the standard models seen in previous papers
\citep[eg][]{ADT98}.

\subsection{Density Structure}

For each set of models we examine three densities; n$_\mathrm{H} =
10^2,~10^3$ and $10^4$ \pccm.

For the standard constant density (isochoric) dust-free models the
concept of a hydrogen density is straightforward. However, in
radiation pressure dominated models which have constant gas plus
photon pressure (isobaric models) the concept is not so
straightforward, since the local density varies continually throughout
the models. Since the density sensitive lines like [\sii] $\lambda
\lambda 6717,30$ are most strongly emitted close to the ionization
front, the local $U$ is low. It is in the same region that effectively
all the radiation pressure in the ionizing radiation field
$P_{\mathrm{*}}$ has been accumulated, such that the local pressure
$P_{\mathrm{loc}}= P_{\mathrm{*}}+P_0$.  We therefore have chosen to
set the hydrogen density in the models to be the density in the [\sii]
emission zone, physically near to the ionization front, but where
there are still sufficient ionizing photons to ensure that
$n_{\mathrm{HII}}/n_{\mathrm{HI}} \sim 1.0$.

\subsection{Form of the Ionizing Spectrum}

Following the convention for photoionization models of AGN we use a
simple power-law to represent the spectrum of the ionizing source,
with
\begin{equation}
F_\nu \propto \nu^\alpha \quad \nu_\mathrm{min} < \nu < \nu_\mathrm{max}.
\end{equation}
In our models we set $\nu_\mathrm{min} = 5$eV and $\nu_\mathrm{max} =
1000$eV.

We investigate four values of the power-law index $\alpha$, -1.2,
-1.4, -1.7 and -2.0. These encompass the `standard' values usually
adopted.  The factor of proportionality, which determines the total
radiative flux entering the photoionized cloud is set by the
ionization parameter at the front of the cloud, $U_0 = S_\star/(n_0
c)$, where $S_\star$ is the flux of ionizing photons, and the initial
value of the density $n_0$ is set by the initial pressure in the
ionized gas, $P_0$, defined in the previous section.

As the final parameter of the models, we set the value of $U$ defined
at the front of the cloud and hence the total flux entering the cloud.
In the case of the standard isochoric photoionization models the
density at the front of the cloud is fixed. However for the dusty,
radiation pressure dominated, isobaric models, the front end density
$n_0$ is an unknown quantity, because we do not {\it a priori} know
the temperature of the plasma at photoionization equilibrium.  We have
estimated this by assuming a front-end temperature of $T_0 = 20,000$ K
and obtaining the density from $P_0$. This may be an underestimate for
high ionization parameters (which give a high $T_0$) and an
overestimate for the low ionization parameters. However, these errors
will have very little effect on the observed density in the [\sii]
emission zone.

A better ionization parameter to use in this instance would be the
ratio of gas pressure to radiation pressure ($\Xi =
P_\mathrm{rad}/P_\mathrm{gas}$), but we have chosen to adopt $U$ as
the ionization parameter to facilitate comparison of our models with
those published in the literature and to limit confusion.

\section{The Density, Ionization \& Temperature Structure}
\label{sec:structure}

In this section, we investigate the temperature and ionization
structure of the dusty, radiation pressure dominated models. We
explore the dependence of the dusty model structures upon the
parameters and reveal how these differ from the corresponding
structure of the dust-free models.

Figure \ref{fig:a-1.4Z1} demonstrates the one dimensional structure of
the ionized region of a dusty NLR cloud from the front of the cloud to
the point at which hydrogen is 99\% recombined.  The model has an
initial ionization parameter of $U_0 = 10^{-2}$, a metallicity of
$1Z_\odot$, a power-law ionizing SED with an index of $\alpha=-1.4$
and a hydrogen density of n$_\mathrm{H}=1000$\pccm. The distance has
been normalized to allow for easy comparison between the different
models. Fig \ref{fig:a-1.4Z1}a (upper left corner) shows the
temperature and density structure of the nebulae. The inverse
correspondence between the temperature (T$_\mathrm{e}$) and hydrogen
density ($\mathrm{n}_{\mathrm{H}}$) is due to the isobaric nature of
the models. This inverse correspondence is not perfect due to the
addition of the radiation pressure term.  The electron density
(n$_\mathrm{e}$) follows the hydrogen density up until hydrogen begins
to recombine, at which point n$_\mathrm{e}$ drops sharply.  This point
is better seen in figure \ref{fig:a-1.4Z1}b which displays the
ionization structure of helium and the fraction of neutral hydrogen.
 
The classical ionization front is marked by the point at which the
fraction of neutral hydrogen sharply increases and the electron
density drops. Beyond this is a partially ionized region dominated by
X-ray heating and photoionization, including Auger processes. This
region forms for all power-law ionizing spectra. The fully ionized
region is marked by a high temperature $(\sim 2 \times 10^4
\mathrm{K})$ and low density, and is the region in which the higher
ionization species are found. The X-ray ionized region has a lower
temperature $(\sim 1 \times 10^4 \mathrm{K})$, higher density and
contains mainly neutral or singly ionized species which are ionized
and excited by secondary electrons. These electrons are produced as
'knock on' electrons in slowing down the energetic primary electrons
produced by inner-shell photoionization.

The transition from fully ionized to partially ionized is clearly
demarcated in the ionization structure of helium and the heavier
elements.  Figures \ref{fig:a-1.4Z1}c and d show the fractional
ionization structure for two other dominant species: Carbon and
Oxygen. Each ionization species is labelled within the figure. The
change from fully to partially ionized is seen in the transitions of
\heii\ to \hei, \ciii\ to \cii\ and \oiii\ to \oi. The ionization
structure of oxygen is particularly interesting as the narrow peak of
\oii\ is a good marker of the transition point. The fraction of the
\oii\ ion rises once the local ionization parameter reaches low enough
values, but then quickly falls as hydrogen becomes neutral. This is
due to the nearly resonant charge-exchange reaction between \oii\ and
\hi, which locks the oxygen ionization balance to that of hydrogen.

\subsection{Determining Factors of the Model Structure}
 
Of all the parameters we use to define the models, the major
determinant of the NLR cloud structure is the ionization parameter. At
high ionization parameters (figure \ref{fig:U-1}), the relative number
of ionizing photons is increased, which increases the average
temperature and ionization state of the nebulae. With a low ionization
parameter (figure \ref{fig:U-3}), there are not enough photons to
maintain a high ionization state and the majority of the \hii\ region
consists of low ionization species. Both the ionized column and the
distance from the front of the cloud to the ionization front are much
smaller in the low ionization parameter case, due to the smaller
number of ionizing photons impinging on the ionized layer.

The same is seen for the dust-free models. At an ionization parameter
of $U_0=10^{-1}$ the dust-free model with depleted abundance (figure
\ref{fig:ndU-1}) shows a similar high ionization state to the
equivalent dusty model. Likewise, the low ionization parameter
dust-free model (figure \ref{fig:ndU-3}) shows the same low ionization
state as the dusty model with $U_0 =10^{-3}$.

There are several differences between the dusty models and the
dust-free models. Most of these differences are more obvious at the
higher ionization parameter, when dust dominates the opacity and are
discussed in the next section. At the low ionization parameter of
$U_0=10^{-3}$ there are few major differences, but one that is
conspicuous is the larger distance covered by the dust-free
models. This difference predominantly arises due to the dust opacity,
which competes with hydrogen for the ionizing photons, thus reduces
the total ionized column \citep{Netzer93,Kraemer94}.  Even with its
low density, dust is such a dominant opacity source because of two
reasons; it has a broad opacity (see figure \ref{fig:ext}) and it is
able to multiply absorb ionizing photons. This is compared to the
atoms of the gas, which have an ionizing opacity strongly peaked at
their ionization potential, and can only absorb one ionizing photon
before their opacity is altered.

The other factor which causes the disparity in distance is the
different density structure of the two models. For the dust-free
models, their isochoric nature means that the total column depth of
hydrogen is determined only by the distance. In the isobaric dusty
models the increasing density means that the total column of hydrogen
can be reached within a smaller distance.

Even with this effect, the density has little influence upon the
overall temperature and ionization structure of the nebula in either
model.  The density does determine the distance to the recombination
edge from the front of the cloud, but simply because the column depth
is approximately constant for a given $U_0$, ionizing SED and
metallicity. As the distance is normalized within the diagrams, no
effect is seen in the ionization structure. As the ionization
structure does not change, the electron density just scales with the
change in density parameter. The temperature structure varies little
with density, as density only affects collisional de-excitation, which
is a second order effect in the cooling function. Cooling and heating
by dust is a first order effect with density, but this dependence is
through the ionization parameter, which is kept constant when varying
density.

In terms of the ionization and temperature structure, more interesting
are the effects of the variation of either the metallicity or the
ionizing power-law spectrum. Of course, the inclusion of dust into the
models also has a large effect upon both of these. The photoelectric
process on dust is a strong heating mechanism within a photoionized
nebula. The dust can also heat the nebula through the sticking of
preferentially low energy electrons to the grains, thus increasing the
average energy of the gas locally. These processes explain the hotter
temperatures seen in the dusty models in comparison to the dust-free
ones (cf figure \ref{fig:U-1} to figure \ref{fig:ndU-1}). The slope in
T$_\mathrm{e}$ seen with distance (and $U_\mathrm{local}$ in section
\ref{sec:Ulocal}) in the dusty models can also be explained by these
processes. Both these heating processes are dependent upon the local
ionization parameter, being stronger at higher values, and thus
increase closer towards the front of the cloud. Dust also has a
secondary heating effect through the depletion of metals from the gas.

Metallicity strongly affects the temperature structure of a nebula due
to the prominent part that heavy elements play in the cooling
processes. The temperature of a nebula is set by the balance of the
heating and cooling processes which are functions of the temperature
and metallicity of the gas. The heating is dominated by
photoionization, with some contribution from recombination. The
cooling is mainly through emission lines and collisional
losses. Increasing the abundance of metals does increase the heating
by providing greater absorption, with this being especially important
at high $U_0$. However, the cooling is much more strongly affected by
an increase in metals, with a higher abundance meaning stronger
emission lines and greater cross-section for collisions.

Figures \ref{fig:a-1.4Z0.25} and \ref{fig:a-1.4Z4} show two dusty
models with metallicities of $0.25Z_\odot$ and $4Z_\odot$
respectively. The other parameters are set to the fiducial values of
$\log U_0=-2.0$, $\alpha=-1.4$ and n$_\mathrm{H} =1000$\pccm. The
effect of varying the metal abundance upon the temperature structure
is obvious in both diagrams, with the low metallicity model having a
much higher temperature overall than the high metallicity case. The
higher temperature in the low $Z$ model leads to a lower average
density in these isobaric models. This means
that the distance to the ionization front and the end of the model is
increased in the low $Z$ models to obtain the same column depth.

The change in the overall temperature resulting from the metallicity
variation also affects the ionization state somewhat. This is because
the recombination rate is temperature dependent, becoming slower at
high temperatures, and therefore the low $Z$ model displays a larger
fraction of high ionization species in comparison to the model with
$Z=4Z_\odot$.

This region at the inner edge of the cloud which displays these high
ionization species is dominated by the photoionization and line
emission of the metals. As the photon field and temperature decrease,
a second region occurs defined by \heii.  This region can be seen in
the temperature structure of both the low $Z$ model (figure
\ref{fig:a-1.4Z0.25}) and the high $Z$ model (figure
\ref{fig:a-1.4Z4}). In the $Z=0.25Z_\odot$ model the region appears as
a change in slope in the temperature structure, but in the
$Z=4Z_\odot$ model the region can be seen as an obvious dip in the
temperature. This larger dip is due to the increased abundance, and
hence collisional cooling, of the metal ions in this region such as
\oiii\, \siii\ and \neiii. These are strong coolants in the visible,
as can be seen by the strength of the [\oiii]$\lambda5007$ line, and
counteract the heating due to photoionization.  The effects of this
temperature drop can be seen in the density as well as in the
ionization structure of carbon and oxygen due to the increase in the
recombination rate. The \oii\ peak is broad at high $Z$ compared to
the low $Z$ model as the low temperature leads to the early
recombination of \oiii.

After the helium dominated region there is a sharp transition zone in
which hydrogen recombines and the \heii\ fraction falls. This region
is dominated by the H photoionization and line cooling. The
equilibrium temperature in this region is $\sim 10^4$ K, and hence
there is a sharp drop seen in the temperature structure of the hot low
$Z$ model and a rise seen in the cool high $Z$ model structure. The
region is very narrow due to the rapid absorption of the H ionizing
photons once the neutral ion fraction becomes significant.  The final
region is the partially ionized zone discussed before, dominated by
X-ray photoionization and secondary electron ionization. The
temperature equilibrium in this region is reasonably insensitive to
metallicity, thus the similarity in appearance between the two models.

The change in structure due to metallicity variation has observational
consequences. Apart from temperature sensitive diagnostic ratios like
[\oiii]$\lambda4363$/[\oiii]$\lambda5007$, there are diagnostics
sensitive to the increased ionization state such as
\ciii]$\lambda1909$/\cii]$\lambda2326$. These, along with abundance
sensitive diagnostics like [\nii]$\lambda6583$/H$\alpha$, can be used
to estimate the metallicity of NLR regions. These diagnostics are
discussed in detail for both the dusty and the dust-free models in the
following paper of this series (Paper 2).

The index of the ionizing power-law directly affects the
temperature. As the power-law becomes flatter the, average energy of
the ionizing photons increases. This leads to an increase of the
average energy per photoionization and thus an overall temperature
increase in the nebula.

Figures \ref{fig:a-1.2Z1} and \ref{fig:a-2.0Z1} show two dusty models
with a flatter ($\alpha=-1.2$) and steeper ($\alpha=-2.0$) ionizing
power-law spectrum respectively. The other parameters are at the
fiducial values of $U_0 = 10^{-2}$, $Z=1Z_\odot$ and
n$_\mathrm{H}=1000$\pccm. The temperature is clearly hotter in the
$\alpha=-1.2$ model than the $\alpha=-2.0$ one. The ionization state
within the $\alpha=-1.2$ model is also clearly higher than the steeper
model. The reason for this is twofold. The flatter model has a hotter
temperature than the $\alpha=-2.0$ model and thus a slower
recombination rate. It also has more high energy photons available for
the ionization of the high ionization species. The relative increase
of these high energy photons to the lower energy ionizing photons is
also why the \heiii\ zone appears much larger than the \heii\ zone in
the flatter spectrum model. The reverse is true for the steeper
power-law model ($\alpha=-2.0$) in which the \heii\ zone becomes
dominant due to the increase in \hi\ and \hei\ ionizing photons
relative to the \heii\ ionizing photons. This alters the temperature
structure due to the \heii\ cooling, and the cooling of the ions which
co-exist with \heii\ in this zone.

Beyond the classical ionization front, the extended partially ionized
region arises because of the additional X-rays from the AGN ionizing
source. Thus a change in the slope of the power-law ionizing spectrum
leads to a change in the structure of this region. A flatter power-law
($\alpha = -1.2$) has relatively more X-rays, and hence leads to a
wider partially ionized zone relative to the model with a steeper
power-law ($\alpha = -2.0$).

The effects of a change in the power-law index can also be seen
observationally such as the line diagnostic diagram
\ariii]$\lambda8.98$/\arii]$\lambda6.98$ and
\neiii]$\lambda15.5$/\neii]$\lambda12.8$ discussed in \citet{Groves03}
(Paper2).

\subsection{Ionization Structure with Local Ionization Parameter}
\label{sec:Ulocal}

In order to understand where the emission of bright lines in NLRs is
likely to originate from, we need to look at the cloud structure as a
function of the local ionization parameter.  The local ionization
parameter is a better guide to the ionization of the nebula because
the ionization state is largely determined the ratio of photon density
to gas particle density. Through inspection of the diagrams of the
structure against $U_\mathrm{local}$ it becomes obvious as to why the
dusty, radiation pressure dominated models provide an improvement over
the standard isochoric photoionization models, and how the dusty
models achieve the tight clustering on line diagnostic diagrams that
is characteristic of narrow line regions.

Figure \ref{fig:dUl-2.0} displays the local ionization parameter
structure diagram for a dusty model with solar metallicity
($1Z_\odot$), a power-law index of $\alpha=-1.4$, hydrogen density of
$1000$\pccm\ and an initial ionization parameter of $U_0 =
10^{-2}$. As in the previous structure diagrams this figure is divided
into four parts, showing the temperature (T$_\mathrm{e}$) and density
(n$_\mathrm{H}$, n$_\mathrm{e}$) structure in the upper left quadrant
and the ionization structure of helium, carbon and oxygen in the other
three quadrants. The $x$-axis on all quadrants is
$\log(U_\mathrm{local})$, and therefore is reversed with respect to
the spatial coordinate used in the previous structure diagrams.  A
comparison with figure \ref{fig:a-1.4Z1} reveals the correspondence of
$U_\mathrm{local}$ with distance. Note that $U_\mathrm{local}$ has
been truncated at $10^{-5}$, which does not precisely correspond with
the end of the photoionized model nebula. The transition to the
partially ionized zone occurs at $\log U_\mathrm{local} \sim -4.3$,
but it does not appear as sharp in terms of $\log U_\mathrm{local}$ as
it does in terms of distance.

Figure \ref{fig:ndUl-2.0} shows the dust-free model which has the same
model parameters as the dusty model in figure \ref{fig:dUl-2.0}.  The
density structure is clearly different, with the dust-free model
having a constant hydrogen density, and an electron density determined
entirely by the local state of ionization within the plasma. The
differences between the ionization structure of the dusty and
dust-free models is less obvious, as expected for the low initial
ionization parameter. The high ionization zone is smaller in terms of
$U_\mathrm{local}$, and the ionization zone occupied by \heii, \ciii\
and \oiii\ appears to be broader. The transition to the partially
ionized zone in the dust-free model shows a sharp transition which is
due to the non-varying density of the isochoric model.

The distinctions between the models becomes more apparent at higher
ionization parameters, when dust dominates the local opacity. Figure
\ref{fig:dUl-1.0} shows the dusty model with an initial ionization
parameter of $U_0 = 10^{-1}$. The other parameters are the same as
figure \ref{fig:dUl-2.0}. Above $\log U_\mathrm{local} \sim -2.5$.
Several new ionization species appear because of the higher local
ionization parameter reached in the early parts of the model, and the
electron temperature in these regions is appreciably higher.  Below
$\log U_\mathrm{local} \sim -2.5$ there is little to distinguish the
$U_0 = 10^{-1}$ dusty model from the model with $U_0 = 10^{-2}$
(figure \ref{fig:dUl-2.0}) in either the density or ionization
structure. The two models appear almost interchangeable below this
value of $U_\mathrm{local}$.

The dust-free model shown in figure \ref{fig:ndUl-1.0} has a
distinctly different structure to the dusty model (figure
\ref{fig:dUl-1.0}) and to the low ionization dust-free model (figure
\ref{fig:ndUl-2.0}). The partially ionized zone of the $U_0 =-1.0$
model is similar in most respects to that seen in the $U_0 =-2.0$
model, although the transition occurs at a lower
$U_\mathrm{local}$. The differences become more obvious in the zone
where \oiii\ and \heii\ are the dominant ionization species. These
species occupy a large range of the $\log U_\mathrm{local}$ space,
with the higher ionization species occurring only at the highest
values of $U_\mathrm{local}$. For example, in oxygen all the species
above \oiii\ occur only between $-1.2 \lapprox \log U_\mathrm{local}
\lapprox -1.0$.

In the dust-free models, the atomic species provide the only opacity
source. At high ionization parameter, and with a ionizing spectrum
that extends beyond the \heii\ limit (such as an AGN or power-law
spectrum), the gas is able to maintain a high ionization state. Thus
the high ionization species are the main opacity source and that the
harder ionizing photons are preferentially absorbed to the softer
photons which ionize \hei, \oii\ and \hi. Therefore, the high
ionization region in the dust-free models occupy only a small region
of $U_\mathrm{local}$ space, as the high energy photons represent only
a fraction of the total number of ionizing photons.

In the dusty models, dust, rather than the gas, provides the
predominant opacity source at high ionization parameter.  As its
opacity peaks near 13.6 eV (see figure \ref{fig:ext}), dust removes
the \hi\ ionizing photons preferentially before the harder, \heii\
ionizing photons. Due to this preferential absorption and because dust
opacity is proportional to the number of ionizing photons, the
ionizing spectrum hardens with decreasing $U_\mathrm{local}$. This
results in a much larger extent of the $U_\mathrm{local}$ space being
occupied by the high ionization species. However, the only differences
in the total column depths of the high ionization species in the two
models arise due to the difference in recombination rates because of
the addition of photoelectric heating to the electron temperature and
the small addition of dust opacity to the high energy photons.

The differences between the dusty and dust-free models become even
more prominent at the very high ionization parameter of $\log U_0
=0.0$. The dusty model in figure \ref{fig:dUl0.0} simply extends the
structure seen in the lower $U_0$ dusty models to higher values.  For
values of $U_\mathrm{local}$ below $-1.0$, the $U_0 =10^{0}$ model
structure s almost identical to the dusty model at $U_0 = 10^{-1}$ in
figure \ref{fig:dUl-1.0}.  Above this value of $U_\mathrm{local}$,
coronal species start to appear due to the high ionization parameter.
By comparison, the ionization structure of the dust-free model at
$\log U_0 =0.0$ (figure \ref{fig:ndUl0.0}) is clearly distinguishable
from both the dusty model (figure \ref{fig:dUl0.0}) and the lower
ionization parameter dust-free model (figure
\ref{fig:ndUl-1.0}). Again, the structure diagram is dominated by the
\heii\ and \oiii\ ionization zone, but unlike the $U_0 = 10^{-1}$
model, in the $U_0 = 10^{0}$ dust-free model this zone extends almost
through the entire model. \heiii\ and the high ionization oxygen
species occur only between $-0.2 \lapprox \log U_\mathrm{local}
\lapprox 0.0$. As in the other models, the partially ionized zone
appears only when $\log U_\mathrm{local} \lapprox -5.0$.

A large part of the distinction between the structure of the two
models arises because of the dust opacity. As the dust extinction is
broad (see figure \ref{fig:ext}) and dominates the opacity at high
$U_0$, the ionizing spectrum in the dusty models alters in a different
manner with $U_\mathrm{local}$ to the dust-free case. Figure
\ref{fig:locspec} shows the variation in ionizing spectrum with
$U_\mathrm{local}$ in both the dusty and dust-free models with $\log
U_0 = -1.0$, $Z=1Z_\odot$, $\alpha=-1.4$, and
n$_\mathrm{H}=1000$\pccm.  The curves are at $\log U_\mathrm{local}
=-1.0,-2.0, -3.0$ and $-4.0$.

At $\log U_\mathrm{local} =-1.0$ (top solid curve), the dusty model
has less overall flux due to the difference in density structure
between the two models.  Below $U_\mathrm{local} =10^{-1}$ the most
obvious difference is the way in which the dust removes photons across
the spectrum, especially visible below the Lyman limit and between the
H absorption and He absorption ($\sim 30$ and 54 eV). This
demonstrates clearly the "hardening" of the spectrum with
$U_\mathrm{local}$ by dust discussed above, while in the dust-free
case the heavier ions (especially He) have remove the majority of the
high energy photons above 50 eV. In the dust case the H edge (13.6 eV)
does not become obvious until $U_\mathrm{local} \sim 10^{-3}$,
demonstrating the competition between dust and H at high $U$.

As discussed in DG02, the dust not only dominates the local opacity at
high ionization parameter ($U_0 \gapprox 10^{-2}$), but also couples
the radiation pressure to the gas pressure. At these values of $U_0$
the radiation pressure is comparable to or greater than the gas
pressure, and hence the nebula structure is determined by the
radiation pressure on dust. An increase in ionization parameter leads
to an increase in dust opacity and radiation pressure, which in turn
leads to a corresponding change in the density. Thus, the
$U_\mathrm{local}$ structure becomes independent of the initial
ionization parameter. Because the $U_\mathrm{local}$ parameter
determines the local emission spectrum in a nebula, this coupling
leads to an emission spectrum which is almost independent of the
initial ionization parameter.  This self-regulatory mechanism lies at
the core of the dusty photoionization paradigm.

\section{Spectral Line Intensities}\label{sec:spectables}

In Tables \ref{tab:dUV0.25} to \ref{tab:dIR4} we present the
intensities of a wide range of emission lines with respect to H$\beta$
from our dusty models. In tables \ref{tab:ndUV0.25} to \ref{tab:ndIR4}
we present the same lines for the classical, dust-free isochoric
models, which have been artificially depleted to the dusty models gas
abundances. In tables \ref{tab:ndunDUV0.25} to \ref{tab:ndunDIR4} we
present the undepleted version of the classical models. These lines
cover a range from the Far UV to the Mid IR and encompass both
temperature and density sensitive diagnostics. Also included are lines
which are prime diagnostics to distinguish shock excitation from
photoionization, as well as lines that can distinguish \hii\ regions
from AGN. Most lines are fairly insensitive to density effects and
hence we have limited ourselves to models with a Hydrogen density of
n$_\mathrm{H} \sim $ 1000 \pccm.  The effects of density variations
will be explored further in the following paper. To minimize the size
of the tables themselves we have only given the lines in steps of -1
dex over $\log U_0$. A full line list from all dusty and dust-free
models are available at
\verb|http://www.mso.anu.edu.au/~bgroves/linedata|.

For both the dusty and dust-free models we present the line ratios in
three wavelength sets: UV, Visible and IR. Each set is comprised of 5
tables, one for each metallicity examined in the models. The tables
themselves are separated into four sections, for the different
power-law indices examined, with the index labeled above the
quadrant. Note that the $\log U = 0.0$ dust-free models are not
included within the tables as these models become unphysical at such a
high ionization parameter and with no corresponding change in the
density structure.

One point to note is that in the UV tables we include several
resonance lines like \civ\ $\lambda1550$ and \mgii\
$\lambda2800$. These lines are strong and observed in the NLR of AGN
and need to be evaluated for a reasonable comparison. However
\mapiii\ does not evaluate the multiple scatterings, and hence the
exact transfer, of these lines, but rather estimates the escape
probability (including dust extinction) of these lines. Thus these
lines have a larger uncertainty associated with them compared to the
forbidden lines, like [\nev]$\lambda3426$.

Though the full line sets will be explored in depth using line
diagnostic diagrams in the following papers, there are some general
patterns that are evident from the tables.

The most obvious is that as the ionization parameter is increased we
see a corresponding increase in the ionization state of the
nebula. This is as expected, although there is also a ``stagnation''
in ionization structure around an ionization parameter above $\log U_0
\sim -2$ in the dusty models which is not seen in the dust-free
models. It is above this value of $U_0$ that dust begins to dominate
the absorption and hence the radiation pressure. As discussed in the
previous section, the radiation pressure on dust at this value of $U$
dominates the gas pressure and determines the density, and hence the
local ionization parameter, towards the recombination zone.

The effects of metallicity variation is also apparent.  As the amount
of metals increases from $Z = 0.25 Z_{\odot}$ there is a corresponding
increase in the relative strength of the metal emission lines. However
at $Z \gapprox 2 Z_{\odot}$ the effect of metal cooling begins to
dominate the line emission, and the higher ionization lines becoming
weaker above this value. This corresponds with the obvious dip seen in
the temperature structure in Figure \ref{fig:a-1.4Z4}. A comparison
between the depleted and undepleted models demonstrates the effect of
dust depletion. In most ways it is similar to a change in overall
metallicity due to the effect upon temperature. However there are
variations which are different to metallicity, especially in lines
from heavily depleted elements such as iron. This is due to the sudden
abundance increase in these elements.

The change in the ionizing spectrum has a visible effect on the line
emission. As the incident power-law becomes steeper, with the index
changing from $-1.2$ to $-2.0$, not only do the high ionization lines
become weaker, but the metal line ratios tend to become weaker
relative to H$\beta$ and other Hydrogen lines. With a steeper
power-law, a larger fraction of the ionizing photons lies in the
Hydrogen ionizing band (1 - 1.8 Rydberg), and there are
correspondingly fewer high energy photons available to ionize the high
ionization species. Because Hydrogen sees more ionizing photons at the
same ionization parameter, there is a relative increase in the
Hydrogen line strength.

%
%

\section{Conclusions}

We have described here the implementation of dust and
radiation pressure in the shock \& photoionization code
\mapiii. To demonstrate the applicability of this model and  to gain a
deeper understanding of the physics, we have run a set of models covering
a wide range in density, metallicity, power-law index of the
ionizing spectra and photoionization parameter. We have presented an
examination of the physical and ionization structure of these models and 
the effects that variation of these parameters has upon the structure
and emission lines.
In addition, we have presented a grid of commonly used
UV, Visible and IR diagnostic spectral lines from the models covering
the full parameter space. This data is
intended for use in the comparison with the Narrow Line
Regions of active galaxies and the understanding of the physical
properties of such regions.
In the next paper we will discuss further the resulting emission lines
from the models and compare the grids of both the dusty models and
of the dust-free models with the observational material
for the commonly diagnostic ratios as well as presenting new
FUV and IR diagnostics.

\begin{acknowledgements}
M.D.~acknowledges the support of the Australian National University
and of the Australian Research Council through his ARC Australian Federation
Fellowship and M.D.~and R.S.~acknowledges support through the ARC
Discovery project DP0208445. The authors would like to thank the referee for their incisive and detailed comments which have much improved these papers.
\end{acknowledgements}

\clearpage

\begin{deluxetable}{lr}
\tablecaption{Depletion Factors
\label{tab:depl}}
\tablehead{
\colhead{Element} & \colhead{Depletion\tablenotemark{a}}
}
\startdata
H  & 0.00 \\
He & 0.00 \\
C  & -0.30 \\
N  & -0.22 \\
O  & -0.22 \\
Ne & 0.00 \\
Na & -0.60 \\
Mg & -0.70 \\
Al & -1.70 \\
Si & -1.00 \\
S  & 0.00 \\
Cl & -0.30 \\
Ar & 0.00 \\
Ca & -2.52 \\
Fe & -2.00 \\
Ni & -1.40
\enddata
\tablenotetext{a}{Depletion given as
$\log(X/\mathrm{H})_{\mathrm{gas}}-
\log(X/\mathrm{H})_{\mathrm{ISM}}$}
\end{deluxetable}

\begin{table}
\begin{center}
\caption{Solar Abundance Set\label{tab:solar}}
~\\
\begin{tabular}{|lc|lc|}

\tableline
Element & Abundance\tablenotemark{a} & Element &
Abundance\tablenotemark{a} \\
\tableline
\tableline
H  & ~0.000 & Al & -5.51~ \\
He & -0.987 & Si & -4.49~ \\
C  & -3.61~ & S  & -4.80~ \\
N  & -4.20~ & Cl & -6.72~ \\
O  & -3.31~ & Ar & -5.60~ \\
Ne & -3.92~ & Ca & -5.65~ \\
Na & -5.68~ & Fe & -4.54~ \\
Mg & -4.42~ & Ni & -5.75~ \\
\tableline
\end{tabular}
\end{center}
\tablenotetext{a}{All abundances are logarithmic with respect to Hydrogen}
\end{table}

\singlespace
\clearpage
%
%
\begin{sidewaystable}[!hp]
{\scriptsize
\caption{Dusty UV line ratios wrt H$\beta$, $0.25 \times Z_\odot$\label{tab:dUV0.25}}
\begin{tabular}{c|ccccc|ccccc}
\tableline
\tableline
& & & $\alpha = -1.2$ & & & & & $\alpha = -1.4$ & & \\
\tableline
$\log U_0$ &  0.0 & -1.0 & -2.0 & -3.0 & -4.0 &  0.0 & -1.0 & -2.0 & -3.0 & -4.0 \\ 
\ciii\ $\lambda$977           &  9.426e-02 &  1.797e+00 &  1.133e-02 & 1.192e-03 &  0.000e+00 &  1.489e-01 &  1.667e+00 &  1.736e-02 & 9.137e-04 &  0.000e+00 \\
\niii\ $\lambda$991           &  1.022e-02 &  1.382e-01 &  1.751e-02 & 2.112e-03 &  5.503e-06 &  1.547e-02 &  1.704e-01 &  1.435e-02 & 1.872e-03 &  4.178e-06 \\
\ovi\ $\lambda\lambda$1032,7  &  5.940e+00 &  7.186e-02 &  1.307e-01 & 1.536e-03 &  4.655e-01 &  6.788e+00 &  9.096e-02 &  1.070e-01 & 1.192e-03 &  3.877e-01 \\
\nv\ $\lambda$1240            &  1.880e-01 &  8.478e-02 &  1.565e-02 & 1.438e-02 &  1.303e-04 &  2.464e-01 &  1.059e-01 &  1.874e-02 & 1.122e-02 &  1.302e-04 \\
\siliv+\oiv$]$ $\lambda$1400  &  2.034e-01 &  2.411e+00 &  4.917e+00 & 1.786e+00 &  6.794e-01 &  2.648e-01 &  3.514e+00 &  4.230e+00 & 1.568e+00 &  6.002e-01 \\
\niv$]$ $\lambda$1486         &  3.326e-02 &  3.295e+00 &  1.976e-02 & 2.116e-01 &  3.174e-01 &  5.204e-02 &  2.994e+00 &  2.030e-02 & 1.879e-01 &  2.640e-01 \\
\end{tabular}
}
\end{sidewaystable}
%
%
\begin{sidewaystable}[!hp]
{\scriptsize
\caption{Dusty UV line ratios wrt H$\beta$, $0.5 \times Z_\odot$}
\begin{tabular}{c|ccccc|ccccc}
\tableline
\tableline \vspace{1mm}
& & & $\alpha = -1.2$ & & & & & $\alpha = -1.4$ & & \\
\tableline
$\log U_0$ &  0.0 & -1.0 & -2.0 & -3.0 & -4.0 &  0.0 & -1.0 & -2.0 & -3.0 & -4.0 \\ 
\ciii\ $\lambda$977           &  1.514e-01 &  2.407e+00 &  1.833e-02 & 1.359e-03 &  0.000e+00 &  2.209e-01 &  2.061e+00 &  2.421e-02 & 9.184e-04 &  0.000e+00 \\
\niii\ $\lambda$991           &  2.290e-02 &  3.124e-01 &  2.297e-02 & 2.714e-03 &  8.349e-06 &  3.242e-02 &  3.651e-01 &  1.675e-02 & 2.152e-03 &  6.024e-06 \\
\ovi\ $\lambda\lambda$1032,7  &  7.991e+00 &  1.617e-01 &  1.715e-01 & 7.586e-04 &  4.908e-01 &  8.562e+00 &  1.926e-01 &  1.249e-01 & 1.326e-03 &  4.047e-01 \\
\nv\ $\lambda$1240            &  4.478e-01 &  1.166e-01 &  2.262e-02 & 1.732e-02 &  2.039e-04 &  5.697e-01 &  1.403e-01 &  2.402e-02 & 1.205e-02 &  1.955e-04 \\
\siliv+\oiv$]$ $\lambda$1400  &  2.911e-01 &  3.954e+00 &  6.095e+00 & 2.014e+00 &  1.047e+00 &  3.674e-01 &  5.356e+00 &  4.890e+00 & 1.737e+00 &  9.499e-01 \\
\niv$]$ $\lambda$1486         &  7.855e-02 &  3.657e+00 &  4.372e-02 & 3.606e-01 &  5.638e-01 &  1.180e-01 &  3.402e+00 &  4.006e-02 & 3.118e-01 &  4.633e-01 \\
\end{tabular}
}
\end{sidewaystable}
%
%
\begin{sidewaystable}[!hp]
{\scriptsize
\caption{Dusty UV line ratios wrt H$\beta$, $1 \times Z_\odot$}
\begin{tabular}{c|ccccc|ccccc}
\tableline
\tableline \vspace{1mm}
& & & $\alpha = -1.2$ & & & & & $\alpha = -1.4$ & & \\
\tableline
$\log U_0$ &  0.0 & -1.0 & -2.0 & -3.0 & -4.0 &  0.0 & -1.0 & -2.0 & -3.0 & -4.0 \\ 
\ciii\ $\lambda$977           &  2.303e-01 &  3.045e+00 &  2.324e-02 & 7.618e-04 &  0.000e+00 &  2.912e-01 &  2.302e+00 &  2.328e-02 & 4.115e-04 &  0.000e+00 \\  
\niii\ $\lambda$991           &  5.510e-02 &  7.443e-01 &  2.292e-02 & 1.853e-03 &  8.263e-06 &  6.818e-02 &  7.829e-01 &  1.369e-02 & 1.151e-03 &  5.248e-06 \\  
\ovi\ $\lambda\lambda$1032,7  &  1.041e+01 &  3.844e-01 &  1.712e-01 & 1.294e-03 &  5.153e-01 &  1.023e+01 &  4.095e-01 &  1.021e-01 & 3.850e-04 &  4.148e-01 \\  
\nv\ $\lambda$1240            &  1.116e+00 &  1.631e-01 &  2.565e-02 & 1.098e-02 &  2.142e-04 &  1.335e+00 &  1.762e-01 &  2.158e-02 & 6.216e-03 &  1.850e-04 \\  
\siliv+\oiv$]$ $\lambda$1400  &  4.276e-01 &  6.260e+00 &  6.586e+00 & 2.057e+00 &  1.391e+00 &  5.035e-01 &  7.544e+00 &  4.687e+00 & 1.689e+00 &  1.244e+00 \\  
\niv$]$ $\lambda$1486         &  1.999e-01 &  4.140e+00 &  8.384e-02 & 5.215e-01 &  8.511e-01 &  2.789e-01 &  3.912e+00 &  6.215e-02 & 4.188e-01 &  6.704e-01 \\
\end{tabular}
}
\end{sidewaystable}
%
%
\begin{sidewaystable}[!hp]
{\scriptsize
\caption{Dusty UV line ratios wrt H$\beta$, $2 \times Z_\odot$}
\begin{tabular}{c|ccccc|ccccc}
\tableline
\tableline \vspace{1mm}
& & & $\alpha = -1.2$ & & & & & $\alpha = -1.4$ & & \\
\tableline
$\log U_0$ &  0.0 & -1.0 & -2.0 & -3.0 & -4.0 &  0.0 & -1.0 & -2.0 & -3.0 & -4.0 \\
\ciii\ $\lambda$977           &  3.264e-01 &  3.578e+00 &  1.762e-02 & 1.077e-04 &  0.000e+00 &  3.307e-01 &  2.275e+00 &  1.249e-02 & 4.152e-05 &  0.000e+00 \\  
\niii\ $\lambda$991           &  1.392e-01 &  1.813e+00 &  1.519e-02 & 3.100e-04 &  2.589e-06 &  1.373e-01 &  1.622e+00 &  6.469e-03 & 1.271e-04 &  1.284e-06 \\  
\ovi\ $\lambda\lambda$1032,7  &  1.308e+01 &  9.335e-01 &  1.138e-01 & 1.754e-04 &  5.202e-01 &  1.148e+01 &  8.397e-01 &  4.839e-02 & 7.478e-05 &  4.128e-01 \\  
\nv\ $\lambda$1240            &  2.904e+00 &  2.193e-01 &  1.847e-02 & 2.010e-03 &  7.433e-05 &  3.145e+00 &  1.987e-01 &  1.112e-02 & 8.352e-04 &  5.320e-05 \\  
\siliv+\oiv$]$ $\lambda$1400  &  6.241e-01 &  9.348e+00 &  5.774e+00 & 1.761e+00 &  1.179e+00 &  6.524e-01 &  9.470e+00 &  3.655e+00 & 1.396e+00 &  9.760e-01 \\  
\niv$]$ $\lambda$1486         &  5.446e-01 &  4.853e+00 &  1.141e-01 & 5.237e-01 &  8.383e-01 &  6.781e-01 &  4.575e+00 &  6.155e-02 & 3.584e-01 &  6.054e-01 \\
\end{tabular}
}
\end{sidewaystable}
%
%
\begin{sidewaystable}[!hp]
{\scriptsize
\caption{Dusty UV line ratios wrt H$\beta$, $4 \times Z_\odot$}
\begin{tabular}{c|ccccc|ccccc}
\tableline
\tableline \vspace{1mm}
& & & $\alpha = -1.2$ & & & & & $\alpha = -1.4$ & & \\
\tableline
$\log U_0$ &  0.0 & -1.0 & -2.0 & -3.0 & -4.0 &  0.0 & -1.0 & -2.0 & -3.0 & -4.0 \\ 
\ciii\ $\lambda$977           &  3.989e-01 &  3.774e+00 &  8.384e-03 & 2.670e-06 &  0.000e+00 &  3.090e-01 &  1.937e+00 &  3.532e-03 & 0.000e+00 &  0.000e+00 \\ 
\niii\ $\lambda$991           &  3.226e-01 &  4.151e+00 &  5.903e-03 & 9.254e-06 &  0.000e+00 &  2.379e-01 &  3.041e+00 &  1.280e-03 & 2.431e-06 &  0.000e+00 \\ 
\ovi\ $\lambda\lambda$1032,7  &  1.563e+01 &  2.128e+00 &  4.438e-02 & 1.188e-05 &  4.666e-01 &  1.202e+01 &  1.562e+00 &  9.596e-03 & 2.811e-06 &  3.519e-01 \\ 
\nv\ $\lambda$1240            &  7.359e+00 &  2.670e-01 &  3.780e-02 & 8.189e-05 &  3.849e-06 &  7.016e+00 &  1.953e-01 &  2.875e-03 & 2.430e-05 &  1.632e-06 \\ 
\siliv+\oiv$]$ $\lambda$1400  &  8.556e-01 &  1.276e+01 &  4.461e+00 & 1.524e+00 &  3.737e-01 &  7.630e-01 &  1.054e+01 &  2.835e+00 & 1.209e+00 &  2.732e-01 \\ 
\niv$]$ $\lambda$1486         &  1.461e+00 &  5.932e+00 &  9.893e-02 & 2.722e-01 &  3.053e-01 &  1.571e+00 &  5.463e+00 &  3.270e-02 & 1.314e-01 &  1.898e-01 \\
\end{tabular}
}
\end{sidewaystable}

%
%
\begin{sidewaystable}[!hp]
\centering
{\scriptsize
\caption{Dusty visible line ratios wrt H$\beta$, $0.25 \times Z_\odot$}
\begin{tabular}{c|ccccc|ccccc}
\tableline
\tableline
& & & $\alpha = -1.2$ & & & & & $\alpha = -1.4$ & & \\
\tableline
$\log U_0$ &  0.0 & -1.0 & -2.0 & -3.0 & -4.0 & 0.0 & -1.0 & -2.0 & -3.0 & -4.0 \\ 
$[$\oii$]$ $\lambda\lambda$3727,9     &  3.694e$-$1 &  3.645e$-$1 &  5.551e$-$1 & 1.873e$+$0 &  1.359e$+$0 &  3.100e$-$1 &  3.130e$-$1 &  5.405e$-$1 & 1.918e$+$0 &  1.390e$+$0 \\ 
$[$\sii$]$ $\lambda\lambda$4067,76    &  6.074e$-$2 &  7.862e$-$2 &  1.035e$-$1 & 1.847e$-$2 &  2.857e$-$1 &  8.713e$-$2 &  1.148e$-$1 &  1.100e$-$1 & 1.830e$-$2 &  2.404e$-$1 \\ 
$[$\oiii$]$ $\lambda$4363             &  3.056e$-$2 &  2.853e$-$2 &  2.840e$-$2 & 7.306e$-$2 &  1.538e$-$1 &  2.552e$-$2 &  2.269e$-$2 &  2.272e$-$2 & 6.259e$-$2 &  1.390e$-$1 \\ 
\heii\ $\lambda$4686                  &  4.464e$-$1 &  4.258e$-$1 &  3.104e$-$1 & 2.060e$-$1 &  2.028e$-$1 &  4.314e$-$1 &  3.860e$-$1 &  2.599e$-$1 & 1.755e$-$1 &  1.704e$-$1 \\ 
$[$\oiii$]$ $\lambda$4959             &  6.474e$-$1 &  8.425e$-$1 &  1.382e$+$0 & 4.610e$-$1 &  4.940e$-$1 &  9.104e$-$1 &  1.238e$+$0 &  1.556e$+$0 & 4.701e$-$1 &  4.150e$-$1 \\ 
$[$\oiii$]$ $\lambda$5007             &  1.865e$+$0 &  2.427e$+$0 &  3.981e$+$0 & 1.328e$+$0 &  1.788e$-$1 &  2.622e$+$0 &  3.564e$+$0 &  4.481e$+$0 & 1.354e$+$0 &  1.525e$-$1 \\
\end{tabular}
}
\end{sidewaystable}
%
%
\begin{sidewaystable}[!hp]
\centering
{\scriptsize
\caption{Dusty visible line ratios wrt H$\beta$, $0.5 \times Z_\odot$}
\begin{tabular}{c|ccccc|ccccc}
\tableline
\tableline \vspace{1mm}
& & & $\alpha = -1.2$ & & & & & $\alpha = -1.4$ & & \\
\tableline
$\log U_0$ &  0.0 & -1.0 & -2.0 & -3.0 & -4.0 & 0.0 & -1.0 & -2.0 & -3.0 & -4.0 \\ 
$[$\oii$]$ $\lambda\lambda$3727,9     &  9.318e$-$1 &  9.107e$-$1 &  1.166e$+$0 & 3.477e$+$0 &  2.529e$+$0 &  7.757e$-$1 &  7.618e$-$1 &  1.098e$+$0 & 3.473e$+$0 &  2.549e$+$0 \\ 
$[$\sii$]$ $\lambda\lambda$4067,76    &  9.127e$-$2 &  1.127e$-$1 &  1.536e$-$1 & 2.766e$-$2 &  6.311e$-$1 &  1.260e$-$1 &  1.571e$-$1 &  1.487e$-$1 & 2.553e$-$2 &  5.259e$-$1 \\ 
$[$\oiii$]$ $\lambda$4363             &  6.896e$-$2 &  6.323e$-$2 &  5.391e$-$2 & 1.220e$-$1 &  2.567e$-$1 &  5.682e$-$2 &  4.932e$-$2 &  4.194e$-$2 & 1.058e$-$1 &  2.386e$-$1 \\ 
\heii\ $\lambda$4686                  &  4.965e$-$1 &  4.816e$-$1 &  3.383e$-$1 & 2.135e$-$1 &  4.769e$-$1 &  4.895e$-$1 &  4.480e$-$1 &  2.829e$-$1 & 1.807e$-$1 &  3.989e$-$1 \\ 
$[$\oiii$]$ $\lambda$4959             &  1.097e$+$0 &  1.371e$+$0 &  2.358e$+$0 & 7.849e$-$1 &  1.162e$+$0 &  1.513e$+$0 &  1.966e$+$0 &  2.519e$+$0 & 7.653e$-$1 &  9.716e$-$1 \\ 
$[$\oiii$]$ $\lambda$5007             &  3.159e$+$0 &  3.949e$+$0 &  6.793e$+$0 & 2.261e$+$0 &  4.507e$-$1 &  4.357e$+$0 &  5.663e$+$0 &  7.255e$+$0 & 2.204e$+$0 &  3.795e$-$1 \\
\end{tabular}
}
\end{sidewaystable}
%
%
\begin{sidewaystable}[!hp]
\centering
{\scriptsize
\caption{Dusty visible line ratios wrt H$\beta$, $1 \times Z_\odot$}
\begin{tabular}{c|ccccc|ccccc}
\tableline
\tableline \vspace{1mm}
& & & $\alpha = -1.2$ & & & & & $\alpha = -1.4$ & & \\
\tableline
$\log U_0$ &  0.0 & -1.0 & -2.0 & -3.0 & -4.0 & 0.0 & -1.0 & -2.0 & -3.0 & -4.0 \\ 
$[$\oii$]$ $\lambda\lambda$3727,9     &  2.301e$+$0 &  2.221e$+$0 &  2.296e$+$0 & 5.651e$+$0 &  4.086e$+$0 &  1.897e$+$0 &  1.792e$+$0 &  1.998e$+$0 & 5.296e$+$0 &  3.938e$+$0 \\ 
$[$\sii$]$ $\lambda\lambda$4067,76    &  1.434e$-$1 &  1.697e$-$1 &  1.868e$-$1 & 2.799e$-$2 &  1.325e$+$0 &  1.820e$-$1 &  2.094e$-$1 &  1.505e$-$1 & 2.241e$-$2 &  1.095e$+$0 \\ 
$[$\oiii$]$ $\lambda$4363             &  1.492e$-$1 &  1.348e$-$1 &  9.742e$-$2 & 1.917e$-$1 &  3.997e$-$1 &  1.214e$-$1 &  1.021e$-$1 &  7.229e$-$2 & 1.669e$-$1 &  3.701e$-$1 \\ 
\heii\ $\lambda$4686                  &  5.649e$-$1 &  5.543e$-$1 &  3.746e$-$1 & 2.237e$-$1 &  1.069e$+$0 &  5.694e$-$1 &  5.259e$-$1 &  3.112e$-$1 & 1.880e$-$1 &  8.989e$-$1 \\ 
$[$\oiii$]$ $\lambda$4959             &  1.931e$+$0 &  2.351e$+$0 &  3.629e$+$0 & 1.063e$+$0 &  2.604e$+$0 &  2.541e$+$0 &  3.146e$+$0 &  3.473e$+$0 & 9.510e$-$1 &  2.190e$+$0 \\ 
$[$\oiii$]$ $\lambda$5007             &  5.561e$+$0 &  6.772e$+$0 &  1.045e$+$1 & 3.062e$+$0 &  1.121e$+$0 &  7.320e$+$0 &  9.060e$+$0 &  1.000e$+$1 & 2.739e$+$0 &  9.329e$-$1 \\
\end{tabular}
}
\end{sidewaystable}
%
%
\begin{sidewaystable}[!hp]
\centering
{\scriptsize
\caption{Dusty visible line ratios wrt H$\beta$, $2 \times Z_\odot$}
\begin{tabular}{c|ccccc|ccccc}
\tableline
\tableline \vspace{1mm}
& & & $\alpha = -1.2$ & & & & & $\alpha = -1.4$ & & \\
\tableline
$\log U_0$ &  0.0 & -1.0 & -2.0 & -3.0 & -4.0 & 0.0 & -1.0 & -2.0 & -3.0 & -4.0 \\ 
$[$\oii$]$ $\lambda\lambda$3727,9     &  5.449e$+$0 &  5.101e$+$0 &  3.623e$+$0 & 6.444e$+$0 &  4.366e$+$0 &  4.295e$+$0 &  3.699e$+$0 &  2.678e$+$0 & 5.345e$+$0 &  3.902e$+$0 \\
$[$\sii$]$ $\lambda\lambda$4067,76    &  2.301e$-$1 &  2.532e$-$1 &  1.541e$-$1 & 1.231e$-$2 &  2.628e$+$0 &  2.518e$-$1 &  2.468e$-$1 &  9.423e$-$2 & 7.924e$-$3 &  2.124e$+$0 \\
$[$\oiii$]$ $\lambda$4363             &  3.061e$-$1 &  2.703e$-$1 &  1.582e$-$1 & 2.617e$-$1 &  4.832e$-$1 &  2.424e$-$1 &  1.944e$-$1 &  1.076e$-$1 & 2.190e$-$1 &  4.297e$-$1 \\
\heii\ $\lambda$4686                  &  6.703e$-$1 &  6.598e$-$1 &  4.210e$-$1 & 2.362e$-$1 &  2.235e$+$0 &  6.848e$-$1 &  6.279e$-$1 &  3.468e$-$1 & 1.954e$-$1 &  1.885e$+$0 \\
$[$\oiii$]$ $\lambda$4959             &  3.499e$+$0 &  4.106e$+$0 &  4.419e$+$0 & 8.860e$-$1 &  5.444e$+$0 &  4.239e$+$0 &  4.761e$+$0 &  3.569e$+$0 & 6.949e$-$1 &  4.591e$+$0 \\
$[$\oiii$]$ $\lambda$5007             &  1.008e$+$1 &  1.183e$+$1 &  1.273e$+$1 & 2.552e$+$0 &  2.684e$+$0 &  1.221e$+$1 &  1.371e$+$1 &  1.028e$+$1 & 2.001e$+$0 &  2.136e$+$0 \\
\end{tabular}
}
\end{sidewaystable}
%
%
\begin{sidewaystable}[!hp]
\centering
{\scriptsize
\caption{Dusty visible line ratios wrt H$\beta$, $4 \times Z_\odot$}
\begin{tabular}{c|ccccc|ccccc}
\tableline
\tableline \vspace{1mm}
& & & $\alpha = -1.2$ & & & & & $\alpha = -1.4$ & & \\
\tableline
$\log U_0$ &  0.0 & -1.0 & -2.0 & -3.0 & -4.0 & 0.0 & -1.0 & -2.0 & -3.0 & -4.0 \\ 
$[$\oii$]$ $\lambda\lambda$3727,9     &  1.141e$+$1 &  9.677e$+$0 &  3.704e$+$0 & 3.946e$+$0 &  1.941e$+$0 &  7.802e$+$0 &  5.591e$+$0 &  2.094e$+$0 & 2.764e$+$0 &  1.560e$+$0 \\ 
$[$\sii$]$ $\lambda\lambda$4067,76    &  3.357e$-$1 &  3.199e$-$1 &  8.040e$-$2 & 1.785e$-$3 &  4.961e$+$0 &  2.898e$-$1 &  2.243e$-$1 &  3.365e$-$2 & 9.257e$-$4 &  3.835e$+$0 \\ 
$[$\oiii$]$ $\lambda$4363             &  5.893e$-$1 &  4.950e$-$1 &  2.081e$-$1 & 2.481e$-$1 &  2.867e$-$1 &  4.387e$-$1 &  3.188e$-$1 &  1.195e$-$1 & 1.805e$-$1 &  2.292e$-$1 \\ 
\heii\ $\lambda$4686                  &  8.366e$-$1 &  8.194e$-$1 &  4.792e$-$1 & 2.430e$-$1 &  4.255e$+$0 &  8.552e$-$1 &  7.668e$-$1 &  3.908e$-$1 & 1.963e$-$1 &  3.499e$+$0 \\ 
$[$\oiii$]$ $\lambda$4959             &  5.958e$+$0 &  6.428e$+$0 &  3.874e$+$0 & 3.884e$-$1 &  1.036e$+$1 &  6.180e$+$0 &  5.914e$+$0 &  2.518e$+$0 & 2.694e$-$1 &  8.523e$+$0 \\ 
$[$\oiii$]$ $\lambda$5007             &  1.716e$+$1 &  1.851e$+$1 &  1.116e$+$1 & 1.119e$+$0 &  5.696e$+$0 &  1.780e$+$1 &  1.703e$+$1 &  7.252e$+$0 & 7.759e$-$1 &  3.859e$+$0 \\
\end{tabular}
}
\end{sidewaystable}

\clearpage

%
%
\begin{sidewaystable}[!hp]
\centering
{\scriptsize
\caption{Dusty IR line ratios wrt H$\beta$, $0.25 \times Z_\odot$}
\begin{tabular}{c|ccccc|ccccc}
\tableline
\tableline \vspace{1mm}
& & & $\alpha = -1.2$ & & & & & $\alpha = -1.4$ & & \\
\tableline
 $\log U_0$ &  \phs 0.0 & -1.0 & -2.0 & -3.0 & -4.0 &  \phs 0.0 & -1.0 & -2.0 & -3.0 & -4.0 \\ 
   Br$\alpha$ $\lambda$4.051 & 5.972e$-$2 & 6.043e$-$2 & 6.631e$-$2 & 7.586e$-$2 & 8.485e$-$2  & 5.977e$-$2 & 6.125e$-$2 & 6.714e$-$2 & 7.569e$-$2 & 8.594e$-$2  \\ 
  $[$\arvi$]$ $\lambda$4.530 & 9.960e$-$3 & 9.314e$-$3 & 9.535e$-$4 & 0.000e$+$0 & 0.000e$+$0  & 1.075e$-$2 & 9.023e$-$3 & 5.333e$-$4 & 0.000e$+$0 & 0.000e$+$0  \\ 
  $[$\arii$]$ $\lambda$6.983 & 1.703e$-$2 & 1.544e$-$2 & 1.381e$-$2 & 3.049e$-$2 & 6.315e$-$2  & 1.328e$-$2 & 1.110e$-$2 & 9.799e$-$3 & 2.239e$-$2 & 4.985e$-$2  \\ 
 $[$\ariii$]$ $\lambda$8.983 & 3.863e$-$1 & 3.418e$-$1 & 3.954e$-$2 & 7.316e$-$6 & 0.000e$+$0  & 3.825e$-$1 & 2.920e$-$1 & 1.988e$-$2 & 3.195e$-$6 & 0.000e$+$0  \\ 
  $[$\nevi$]$ $\lambda$7.652 & 1.957e$-$2 & 2.209e$-$2 & 4.496e$-$2 & 5.968e$-$2 & 2.904e$-$2  & 2.019e$-$2 & 2.446e$-$2 & 4.886e$-$2 & 6.242e$-$2 & 2.975e$-$2  \\ 
   $[$\siv$]$ $\lambda$10.44 & 2.219e$-$1 & 2.924e$-$1 & 6.424e$-$1 & 1.125e$-$1 & 8.612e$-$4  & 3.110e$-$1 & 4.493e$-$1 & 7.362e$-$1 & 9.925e$-$2 & 8.658e$-$4  \\
\end{tabular}
}
\end{sidewaystable}
%
%
\begin{sidewaystable}[!hp]
\centering
{\scriptsize
\caption{Dusty IR line ratios wrt H$\beta$, $0.5 \times Z_\odot$}
\begin{tabular}{c|ccccc|ccccc}
\tableline
\tableline \vspace{1mm}
& & & $\alpha = -1.2$ & & & & & $\alpha = -1.4$ & & \\
\tableline
 $\log U_0$ &  \phs 0.0 & -1.0 & -2.0 & -3.0 & -4.0 &  \phs 0.0 & -1.0 & -2.0 & -3.0 & -4.0 \\ 
   Br$\alpha$ $\lambda$4.051 & 6.215e$-$2 & 6.272e$-$2 & 6.838e$-$2 & 7.666e$-$2 & 8.571e$-$2  & 6.236e$-$2 & 6.360e$-$2 & 6.946e$-$2 & 7.670e$-$2 & 8.594e$-$2  \\  
  $[$\arvi$]$ $\lambda$4.530 & 1.682e$-$2 & 1.576e$-$2 & 1.457e$-$3 & 0.000e$+$0 & 0.000e$+$0  & 1.804e$-$2 & 1.529e$-$2 & 7.634e$-$4 & 0.000e$+$0 & 0.000e$+$0  \\  
  $[$\arii$]$ $\lambda$6.983 & 3.996e$-$2 & 3.612e$-$2 & 2.753e$-$2 & 5.292e$-$2 & 1.067e$-$1  & 3.133e$-$2 & 2.610e$-$2 & 1.915e$-$2 & 3.863e$-$2 & 8.663e$-$2  \\  
 $[$\ariii$]$ $\lambda$8.983 & 6.618e$-$1 & 5.912e$-$1 & 6.019e$-$2 & 1.047e$-$5 & 0.000e$+$0  & 6.502e$-$1 & 5.000e$-$1 & 2.809e$-$2 & 4.219e$-$6 & 0.000e$+$0  \\  
  $[$\nevi$]$ $\lambda$7.652 & 4.493e$-$2 & 4.975e$-$2 & 9.418e$-$2 & 1.183e$-$1 & 5.645e$-$2  & 4.660e$-$2 & 5.484e$-$2 & 1.017e$-$1 & 1.226e$-$1 & 5.707e$-$2  \\  
   $[$\siv$]$ $\lambda$10.44 & 4.291e$-$1 & 5.423e$-$1 & 1.228e$+$0 & 1.974e$-$1 & 5.173e$-$3  & 5.852e$-$1 & 8.061e$-$1 & 1.363e$+$0 & 1.675e$-$1 & 1.669e$-$3  \\
\end{tabular}
}
\end{sidewaystable}
%
%
\begin{sidewaystable}[!hp]
\centering
{\scriptsize
\caption{Dusty IR line ratios wrt H$\beta$, $1 \times Z_\odot$}
\begin{tabular}{c|ccccc|ccccc}
\tableline
\tableline \vspace{1mm}
& & & $\alpha = -1.2$ & & & & & $\alpha = -1.4$ & & \\
\tableline
 $\log U_0$ &  \phs 0.0 & -1.0 & -2.0 & -3.0 & -4.0 &  \phs 0.0 & -1.0 & -2.0 & -3.0 & -4.0 \\ 
   Br$\alpha$ $\lambda$4.051 & 6.469e$-$2 & 6.526e$-$2 & 7.137e$-$2 & 7.928e$-$2 & 8.901e$-$2  & 6.522e$-$2 & 6.656e$-$2 & 7.319e$-$2 & 7.949e$-$2 & 9.043e$-$2  \\ 
  $[$\arvi$]$ $\lambda$4.530 & 2.755e$-$2 & 2.564e$-$2 & 1.847e$-$3 & 0.000e$+$0 & 0.000e$+$0  & 2.949e$-$2 & 2.469e$-$2 & 8.795e$-$4 & 0.000e$+$0 & 0.000e$+$0  \\ 
  $[$\arii$]$ $\lambda$6.983 & 8.806e$-$2 & 7.896e$-$2 & 5.211e$-$2 & 8.678e$-$2 & 1.748e$-$1  & 6.945e$-$2 & 5.674e$-$2 & 3.536e$-$2 & 6.360e$-$2 & 1.456e$-$1  \\ 
 $[$\ariii$]$ $\lambda$8.983 & 1.083e$+$0 & 9.621e$-$1 & 7.624e$-$2 & 1.031e$-$5 & 0.000e$+$0  & 1.051e$+$0 & 7.912e$-$1 & 3.177e$-$2 & 3.810e$-$6 & 0.000e$+$0  \\ 
  $[$\nevi$]$ $\lambda$7.652 & 1.011e$-$1 & 1.115e$-$1 & 1.961e$-$1 & 2.266e$-$1 & 1.007e$-$1  & 1.063e$-$1 & 1.234e$-$1 & 2.078e$-$1 & 2.311e$-$1 & 9.869e$-$2  \\ 
   $[$\siv$]$ $\lambda$10.44 & 8.741e$-$1 & 1.083e$+$0 & 2.271e$+$0 & 2.933e$-$1 & 3.749e$-$3  & 1.159e$+$0 & 1.537e$+$0 & 2.379e$+$0 & 2.361e$-$1 & 8.295e$-$3  \\
\end{tabular}
}
\end{sidewaystable}
%
%
\begin{sidewaystable}[!hp]
\centering
{\scriptsize
\caption{Dusty IR line ratios wrt H$\beta$, $2 \times Z_\odot$}
\begin{tabular}{c|ccccc|ccccc}
\tableline
\tableline \vspace{1mm}
& & & $\alpha = -1.2$ & & & & & $\alpha = -1.4$ & & \\
\tableline
 $\log U_0$ &  \phs 0.0 & -1.0 & -2.0 & -3.0 & -4.0 &  \phs 0.0 & -1.0 & -2.0 & -3.0 & -4.0 \\ 
   Br$\alpha$ $\lambda$4.051 & 6.756e$-$2 & 6.846e$-$2 & 7.615e$-$2 & 8.405e$-$2 & 9.463e$-$2 & 6.870e$-$2 & 7.050e$-$2 & 7.904e$-$2 & 8.512e$-$2 & 9.256e$-$2  \\ 
  $[$\arvi$]$ $\lambda$4.530 & 4.396e$-$2 & 4.011e$-$2 & 1.823e$-$3 & 0.000e$+$0 & 0.000e$+$0 & 4.671e$-$2 & 3.764e$-$2 & 7.493e$-$4 & 0.000e$+$0 & 0.000e$+$0  \\ 
  $[$\arii$]$ $\lambda$6.983 & 1.818e$-$1 & 1.598e$-$1 & 9.183e$-$2 & 1.390e$-$1 & 2.878e$-$1 & 1.424e$-$1 & 1.122e$-$1 & 6.092e$-$2 & 1.052e$-$1 & 2.536e$-$1  \\ 
 $[$\ariii$]$ $\lambda$8.983 & 1.695e$+$0 & 1.467e$+$0 & 7.517e$-$2 & 5.999e$-$6 & 0.000e$+$0 & 1.605e$+$0 & 1.142e$+$0 & 2.612e$-$2 & 1.997e$-$6 & 0.000e$+$0  \\ 
  $[$\nevi$]$ $\lambda$7.652 & 2.218e$-$1 & 2.462e$-$1 & 3.926e$-$1 & 3.999e$-$1 & 1.456e$-$1 & 2.381e$-$1 & 2.725e$-$1 & 4.044e$-$1 & 3.962e$-$1 & 1.385e$-$1  \\ 
   $[$\siv$]$ $\lambda$10.44 & 1.811e$+$0 & 2.220e$+$0 & 3.851e$+$0 & 3.282e$-$1 & 9.896e$-$3 & 2.338e$+$0 & 2.973e$+$0 & 3.719e$+$0 & 2.454e$-$1 & 4.877e$-$3  \\
\end{tabular}
}
\end{sidewaystable}
%
%
\begin{sidewaystable}[!hp]
\centering
{\scriptsize
\caption{Dusty IR line ratios wrt H$\beta$, $4 \times Z_\odot$\label{tab:dIR4}}
\begin{tabular}{c|ccccc|ccccc}
\tableline
\tableline \vspace{1mm}
& & & $\alpha = -1.2$ & & & & & $\alpha = -1.4$ & & \\
\tableline
 $\log U_0$ &  \phs 0.0 & -1.0 & -2.0 & -3.0 & -4.0 &  \phs 0.0 & -1.0 & -2.0 & -3.0 & -4.0 \\ 
   Br$\alpha$ $\lambda$4.051 & 7.169e$-$2 & 7.301e$-$2 & 8.374e$-$2 & 9.239e$-$2 & 1.027e$-$1 & 7.391e$-$2 & 7.631e$-$2 & 8.776e$-$2 & 9.360e$-$2 & 1.051e$-$1  \\  
  $[$\arvi$]$ $\lambda$4.530 & 6.792e$-$2 & 5.985e$-$2 & 1.301e$-$3 & 0.000e$+$0 & 0.000e$+$0 & 7.072e$-$2 & 5.320e$-$2 & 4.009e$-$4 & 0.000e$+$0 & 0.000e$+$0  \\  
  $[$\arii$]$ $\lambda$6.983 & 3.553e$-$1 & 3.036e$-$1 & 1.549e$-$1 & 2.373e$-$1 & 4.844e$-$1 & 2.748e$-$1 & 2.065e$-$1 & 1.039e$-$1 & 1.939e$-$1 & 4.311e$-$1  \\  
 $[$\ariii$]$ $\lambda$8.983 & 2.525e$+$0 & 2.072e$+$0 & 5.298e$-$2 & 2.211e$-$6 & 0.000e$+$0 & 2.292e$+$0 & 1.476e$+$0 & 1.322e$-$2 & 0.000e$+$0 & 0.000e$+$0  \\  
  $[$\nevi$]$ $\lambda$7.652 & 4.641e$-$1 & 5.159e$-$1 & 7.255e$-$1 & 5.953e$-$1 & 1.427e$-$1 & 5.009e$-$1 & 5.641e$-$1 & 7.220e$-$1 & 5.679e$-$1 & 1.292e$-$1  \\  
   $[$\siv$]$ $\lambda$10.44 & 3.613e$+$0 & 4.360e$+$0 & 5.736e$+$0 & 2.738e$-$1 & 5.102e$-$3 & 4.453e$+$0 & 5.367e$+$0 & 5.055e$+$0 & 1.945e$-$1 & 3.563e$-$3  \\
\end{tabular}
}
\end{sidewaystable}

\clearpage
%
%
\begin{table}[!hp]
{\scriptsize
\caption{Dust-free UV line ratios wrt H$\beta$, $0.25 \times Z_\odot$\label{tab:ndUV0.25}}
\begin{tabular}{c|cccc|cccc}
\tableline
\tableline
& & & $\alpha = -1.2$ & & & & $\alpha = -1.4$ & \\
\tableline
$\log U_0$ & -1.0 & -2.0 & -3.0 & -4.0 & -1.0 & -2.0 & -3.0 & -4.0 \\ 
\ciii\ $\lambda$977           &  2.749e+00 &  1.865e-02 & 3.438e-03 &  0.000e+00 &  1.722e+00 &  2.431e-02 & 2.662e-03 &  0.000e+00 \\ 
niii\ $\lambda$991            &  3.651e-01 &  1.864e-02 & 4.487e-03 &  3.558e-05 &  3.072e-01 &  1.441e-02 & 4.146e-03 &  2.931e-05 \\ 
\ovi\ $\lambda\lambda$1032,7  &  1.927e-01 &  1.465e-01 & 4.864e-03 &  6.769e-01 &  1.639e-01 &  1.131e-01 & 3.836e-03 &  5.776e-01 \\ 
\nv\ $\lambda$1240            &  2.015e-01 &  1.151e-01 & 4.334e-02 &  6.663e-04 &  1.361e-01 &  9.408e-02 & 3.420e-02 &  6.773e-04 \\ 
\siliv+\oiv$]$ $\lambda$1400  &  7.002e+00 &  5.501e+00 & 2.239e+00 &  7.225e-01 &  7.215e+00 &  4.596e+00 & 1.970e+00 &  6.323e-01 \\ 
\niv$]$ $\lambda$1486         &  2.195e+00 &  4.204e-02 & 1.971e-01 &  4.149e-01 &  1.776e+00 &  4.079e-02 & 1.742e-01 &  3.609e-01 \\
\end{tabular}
}
\end{table}

%
%
\begin{table}[!hp]
{\scriptsize
\caption{Dust-free UV line ratios wrt H$\beta$, $0.5 \times Z_\odot$}
\begin{tabular}{c|cccc|cccc}
\tableline
\tableline \vspace{1mm}
& & & $\alpha = -1.2$ & & & & $\alpha = -1.4$ & \\
\tableline
$\log U_0$ & -1.0 & -2.0 & -3.0 & -4.0 & -1.0 & -2.0 & -3.0 & -4.0 \\ 
\ciii\ $\lambda$977           &  3.498e+00 &  2.753e-02 & 4.493e-03 &  0.000e+00 &  2.002e+00 &  2.986e-02 & 3.165e-03 &  0.000e+00 \\ 
niii\ $\lambda$991            &  4.638e-01 &  2.492e-02 & 6.457e-03 &  5.530e-05 &  3.573e-01 &  1.726e-02 & 5.474e-03 &  4.368e-05 \\ 
\ovi\ $\lambda\lambda$1032,7  &  2.435e-01 &  1.955e-01 & 6.359e-03 &  7.035e-01 &  1.890e-01 &  1.352e-01 & 4.577e-03 &  5.985e-01 \\ 
\nv\ $\lambda$1240            &  2.229e-01 &  3.183e-02 & 6.017e-02 &  1.065e-03 &  1.394e-01 &  2.999e-02 & 4.359e-02 &  1.045e-03 \\ 
\siliv+\oiv$]$ $\lambda$1400  &  7.852e+00 &  6.973e+00 & 2.615e+00 &  1.125e+00 &  7.153e+00 &  5.354e+00 & 2.264e+00 &  1.001e+00 \\ 
\niv$]$ $\lambda$1486         &  2.242e+00 &  5.727e-02 & 3.361e-01 &  7.404e-01 &  1.777e+00 &  4.841e-02 & 2.859e-01 &  6.382e-01 \\
\end{tabular}
}
\end{table}
%
%
\begin{table}[!hp]
{\scriptsize
\caption{Dust-free UV line ratios wrt H$\beta$, $1 \times Z_\odot$}
\begin{tabular}{c|cccc|cccc}
\tableline
\tableline \vspace{1mm}
& & & $\alpha = -1.2$ & & & & $\alpha = -1.4$ & \\
\tableline
$\log U_0$ & -1.0 & -2.0 & -3.0 & -4.0 & -1.0 & -2.0 & -3.0 & -4.0 \\  
\ciii\ $\lambda$977           &  3.661e+00 &  2.639e-02 & 3.461e-03 &  1.016e-06 &  1.899e+00 &  2.192e-02 & 1.961e-03 &  0.000e+00 \\ 
niii\ $\lambda$991            &  8.712e-01 &  2.473e-02 & 5.758e-03 &  5.866e-05 &  6.129e-01 &  1.472e-02 & 3.898e-03 &  4.149e-05 \\ 
\ovi\ $\lambda\lambda$1032,7  &  4.476e-01 &  1.939e-01 & 3.295e-03 &  7.462e-01 &  3.158e-01 &  1.152e-01 & 2.027e-03 &  6.312e-01 \\ 
\nv\ $\lambda$1240            &  2.003e-01 &  3.040e-02 & 5.194e-02 &  1.205e-03 &  1.176e-01 &  9.471e-02 & 3.065e-02 &  1.078e-03 \\ 
\siliv+\oiv$]$ $\lambda$1400  &  6.962e+00 &  7.294e+00 & 2.752e+00 &  1.513e+00 &  5.844e+00 &  4.940e+00 & 2.231e+00 &  1.331e+00 \\ 
\niv$]$ $\lambda$1486         &  2.237e+00 &  9.718e-02 & 4.827e-01 &  1.156e+00 &  1.746e+00 &  6.686e-02 & 3.731e-01 &  9.650e-01 \\
\end{tabular}
}
\end{table}
%
%
\begin{table}[!hp]
{\scriptsize
\caption{Dust-free UV line ratios wrt H$\beta$, $2 \times Z_\odot$}
\begin{tabular}{c|cccc|cccc}
\tableline
\tableline \vspace{1mm}
& & & $\alpha = -1.2$ & & & & $\alpha = -1.4$ & \\
\tableline
$\log U_0$ & -1.0 & -2.0 & -3.0 & -4.0 &  -1.0 & -2.0 & -3.0 & -4.0 \\ 
\ciii\ $\lambda$977           &  3.207e+00 &  1.396e-02 & 9.397e-04 &  0.000e+00 &  1.508e+00 &  9.954e-03 & 3.778e-04 &  0.000e+00 \\ 
niii\ $\lambda$991            &  1.552e+00 &  1.777e-02 & 1.709e-03 &  2.239e-05 &  1.001e+00 &  8.818e-03 & 7.787e-04 &  1.239e-05 \\ 
\ovi\ $\lambda\lambda$1032,7  &  7.900e-01 &  1.394e-01 & 1.727e-03 &  7.968e-01 &  5.103e-01 &  6.902e-02 & 1.063e-03 &  6.659e-01 \\ 
\nv\ $\lambda$1240            &  1.552e-01 &  1.815e-02 & 1.675e-02 &  5.255e-04 &  8.899e-02 &  1.123e-02 & 7.073e-03 &  3.836e-04 \\ 
\siliv+\oiv$]$ $\lambda$1400  &  5.910e+00 &  6.053e+00 & 2.296e+00 &  1.388e+00 &  4.753e+00 &  3.749e+00 & 1.781e+00 &  1.112e+00 \\ 
\niv$]$ $\lambda$1486         &  2.200e+00 &  1.190e-01 & 4.886e-01 &  1.272e+00 &  1.697e+00 &  6.709e-02 & 3.150e-01 &  9.668e-01 \\
\end{tabular}
}
\end{table}
%
%
\begin{table}[!hp]
{\scriptsize
\caption{Dust-free UV line ratios wrt H$\beta$, $4 \times Z_\odot$}
\begin{tabular}{c|cccc|cccc}
\tableline
\tableline \vspace{1mm}
& & & $\alpha = -1.2$ & & & & $\alpha = -1.4$ & \\
\tableline
$\log U_0$ & -1.0 & -2.0 & -3.0 & -4.0 &  -1.0 & -2.0 & -3.0 & -4.0 \\
\ciii\ $\lambda$977           &  2.431e+00 &  7.427e-03 & 8.219e-05 &  0.000e+00 &  1.030e+00 &  4.701e-03 & 2.274e-05 &  0.000e+00 \\ 
niii\ $\lambda$991            &  2.597e+00 &  1.015e-02 & 1.385e-04 &  1.739e-06 &  1.537e+00 &  3.972e-03 & 4.067e-05 &  0.000e+00 \\ 
\ovi\ $\lambda\lambda$1032,7  &  1.322e+00 &  7.996e-02 & 3.349e-04 &  8.311e-01 &  7.837e-01 &  3.123e-02 & 1.219e-04 &  6.852e-01 \\ 
\nv\ $\lambda$1240            &  1.151e-01 &  9.989e-03 & 1.811e-03 &  4.939e-05 &  6.620e-02 &  5.194e-03 & 5.309e-04 &  2.973e-05 \\ 
\siliv+\oiv$]$ $\lambda$1400  &  5.797e+00 &  4.701e+00 & 1.987e+00 &  5.891e-01 &  4.356e+00 &  2.799e+00 & 1.591e+00 &  4.009e-01 \\ 
\niv$]$ $\lambda$1486         &  2.171e+00 &  1.303e-01 & 3.195e-01 &  6.711e-01 &  1.661e+00 &  6.088e-02 & 1.538e-01 &  4.373e-01 \\
\end{tabular}
}
\end{table}

\clearpage
%
%
\begin{table}[!hp]
{\scriptsize
\caption{Dust-free visible line ratios wrt H$\beta$, $0.25 \times Z_\odot$}
\begin{tabular}{c|cccc|cccc}
\tableline
\tableline
& & $\alpha = -1.2$ & & & & $\alpha = -1.4$ & & \\
\tableline
$\log U_0$ & -1.0 & -2.0 & -3.0 & -4.0 & -1.0 & -2.0 & -3.0 & -4.0 \\ 
$[$\oii$]$ $\lambda\lambda$3727,9     &  5.011e$-$2 &  3.585e$-$1 & 1.614e$+$0 &  1.669e$+$0 &  4.626e$-$2 &  3.373e$-$1 & 1.632e$+$0 &  1.718e$+$0 \\ 
$[$\sii$]$ $\lambda\lambda$4067,76    &  2.874e$-$1 &  1.446e$-$1 & 3.176e$-$2 &  4.770e$-$4 &  2.552e$-$1 &  1.445e$-$1 & 3.202e$-$2 &  4.822e$-$4 \\ 
$[$\oiii$]$ $\lambda$4363             &  2.804e$-$3 &  1.457e$-$2 & 5.497e$-$2 &  1.326e$-$1 &  2.348e$-$3 &  1.226e$-$2 & 4.736e$-$2 &  1.202e$-$1 \\ 
\heii\ $\lambda$4686                  &  2.713e$-$1 &  2.864e$-$1 & 2.346e$-$1 &  1.038e$-$1 &  2.222e$-$1 &  2.337e$-$1 & 1.976e$-$1 &  8.861e$-$2 \\ 
$[$\oiii$]$ $\lambda$4959             &  2.455e$+$0 &  1.815e$+$0 & 7.100e$-$1 &  2.168e$-$2 &  2.545e$+$0 &  1.947e$+$0 & 7.370e$-$1 &  2.250e$-$2 \\ 
$[$\oiii$]$ $\lambda$5007             &  7.071e$+$0 &  5.228e$+$0 & 2.045e$+$0 &  6.244e$-$2 &  7.330e$+$0 &  5.608e$+$0 & 2.123e$+$0 &  6.479e$-$2 \\
\end{tabular}
}
\end{table}
%
%
\begin{table}[!hp]
{\scriptsize
\caption{Dust-free visible line ratios wrt H$\beta$, $0.5 \times Z_\odot$}
\begin{tabular}{c|cccc|cccc}
\tableline
\tableline \vspace{1mm}
& & $\alpha = -1.2$ & & & & $\alpha = -1.4$ & & \\
\tableline
$\log U_0$ & -1.0 & -2.0 & -3.0 & -4.0 & -1.0 & -2.0 & -3.0 & -4.0 \\ 
$[$\oii$]$ $\lambda\lambda$3727,9     &  8.501e$-$2 &  6.627e$-$1 & 2.952e$+$0 &  3.064e$+$0 &  7.674e$-$2 &  5.984e$-$1 & 2.895e$+$0 &  3.115e$+$0 \\ 
$[$\sii$]$ $\lambda\lambda$4067,76    &  3.444e$-$1 &  2.223e$-$1 & 5.074e$-$2 &  9.102e$-$4 &  2.688e$-$1 &  1.956e$-$1 & 4.807e$-$2 &  9.251e$-$4 \\ 
$[$\oiii$]$ $\lambda$4363             &  4.873e$-$3 &  2.599e$-$2 & 9.310e$-$2 &  2.191e$-$1 &  3.938e$-$3 &  2.131e$-$2 & 7.977e$-$2 &  2.039e$-$1 \\ 
\heii\ $\lambda$4686                  &  2.825e$-$1 &  2.940e$-$1 & 2.414e$-$1 &  1.082e$-$1 &  2.272e$-$1 &  2.391e$-$1 & 2.031e$-$1 &  9.212e$-$2 \\ 
$[$\oiii$]$ $\lambda$4959             &  3.781e$+$0 &  3.226e$+$0 & 1.269e$+$0 &  3.977e$-$2 &  3.604e$+$0 &  3.205e$+$0 & 1.272e$+$0 &  4.062e$-$2 \\ 
$[$\oiii$]$ $\lambda$5007             &  1.089e$+$1 &  9.290e$+$0 & 3.654e$+$0 &  1.146e$-$1 &  1.038e$+$1 &  9.231e$+$0 & 3.664e$+$0 &  1.170e$-$1 \\
\end{tabular}
}
\end{table}
%
%
\begin{table}[!hp]
{\scriptsize
\caption{Dust-free visible line ratios wrt H$\beta$, $1 \times Z_\odot$}
\begin{tabular}{c|cccc|cccc}
\tableline
\tableline \vspace{1mm}
& & $\alpha = -1.2$ & & & & $\alpha = -1.4$ & & \\
\tableline
$\log U_0$ & -1.0 & -2.0 & -3.0 & -4.0 & -1.0 & -2.0 & -3.0 & -4.0 \\ 
$[$\oii$]$ $\lambda\lambda$3727,9     &  1.245e$-$1 &  1.083e$+$0 & 4.689e$+$0 &  4.952e$+$0 &  1.104e$-$1 &  9.044e$-$1 & 4.268e$+$0 &  4.853e$+$0 \\ 
$[$\sii$]$ $\lambda\lambda$4067,76    &  3.137e$-$1 &  2.521e$-$1 & 5.913e$-$2 &  1.624e$-$3 &  2.121e$-$1 &  1.821e$-$1 & 4.833e$-$2 &  1.603e$-$3 \\ 
$[$\oiii$]$ $\lambda$4363             &  7.141e$-$3 &  4.279e$-$2 & 1.448e$-$1 &  3.392e$-$1 &  5.489e$-$3 &  3.339e$-$2 & 1.208e$-$1 &  3.175e$-$1 \\ 
\heii\ $\lambda$4686                  &  2.882e$-$1 &  3.027e$-$1 & 2.539e$-$1 &  1.156e$-$1 &  2.286e$-$1 &  2.450e$-$1 & 2.133e$-$1 &  9.798e$-$2 \\ 
$[$\oiii$]$ $\lambda$4959             &  4.903e$+$0 &  4.857e$+$0 & 1.924e$+$0 &  6.366e$-$2 &  4.244e$+$0 &  4.263e$+$0 & 1.779e$+$0 &  6.250e$-$2 \\ 
$[$\oiii$]$ $\lambda$5007             &  1.412e$+$1 &  1.399e$+$1 & 5.542e$+$0 &  1.833e$-$1 &  1.222e$+$1 &  1.228e$+$1 & 5.124e$+$0 &  1.800e$-$1 \\
\end{tabular}
}
\end{table}
%
%
\begin{table}[!hp]
{\scriptsize
\caption{Dust-free visible line ratios wrt H$\beta$, $2 \times Z_\odot$}
\begin{tabular}{c|cccc|cccc}
\tableline
\tableline \vspace{1mm}
& & $\alpha = -1.2$ & & & & $\alpha = -1.4$ & & \\
\tableline
$\log U_0$ & -1.0 & -2.0 & -3.0 & -4.0 & -1.0 & -2.0 & -3.0 & -4.0 \\ 
$[$\oii$]$ $\lambda\lambda$3727,9     &  1.278e$-$1 &  1.328e$+$0 & 5.194e$+$0 &  5.537e$+$0 &  1.087e$-$1 &  9.700e$-$1 & 4.101e$+$0 &  4.940e$+$0 \\ 
$[$\sii$]$ $\lambda\lambda$4067,76    &  2.141e$-$1 &  1.850e$-$1 & 3.430e$-$2 &  0.000e$+$0 &  1.248e$-$1 &  1.089e$-$1 & 2.208e$-$2 &  1.976e$-$3 \\ 
$[$\oiii$]$ $\lambda$4363             &  7.140e$-$3 &  6.006e$-$2 & 1.878e$-$1 &  4.214e$-$1 &  4.975e$-$3 &  4.305e$-$2 & 1.448e$-$1 &  3.738e$-$1 \\ 
\heii\ $\lambda$4686                  &  2.897e$-$1 &  3.126e$-$1 & 2.749e$-$1 &  1.259e$-$1 &  2.274e$-$1 &  2.526e$-$1 & 2.303e$-$1 &  1.058e$-$1 \\ 
$[$\oiii$]$ $\lambda$4959             &  4.760e$+$0 &  5.492e$+$0 & 2.018e$+$0 &  7.007e$-$2 &  3.650e$+$0 &  4.160e$+$0 & 1.639e$+$0 &  6.333e$-$2 \\ 
$[$\oiii$]$ $\lambda$5007             &  1.371e$+$1 &  1.582e$+$1 & 5.813e$+$0 &  2.018e$-$1 &  1.051e$+$1 &  1.198e$+$1 & 4.721e$+$0 &  1.824e$-$1 \\
\end{tabular}
}
\end{table}
%
%
\begin{table}[!hp]
{\scriptsize
\caption{Dust-free visible line ratios wrt H$\beta$, $4 \times Z_\odot$}
\begin{tabular}{c|cccc|cccc}
\tableline
\tableline \vspace{1mm}
& & $\alpha = -1.2$ & & & & $\alpha = -1.4$ & & \\
\tableline
$\log U_0$ & -1.0 & -2.0 & -3.0 & -4.0 & -1.0 & -2.0 & -3.0 & -4.0 \\ 
$[$\oii$]$ $\lambda\lambda$3727,9     &  4.478e$-$2 &  9.577e$-$1 & 3.242e$+$0 &  2.998e$+$0 &  2.880e$-$2 &  5.294e$-$1 & 2.066e$+$0 &  2.376e$+$0 \\
$[$\sii$]$ $\lambda\lambda$4067,76    &  1.431e$-$1 &  1.064e$-$1 & 8.556e$-$3 &  1.691e$-$3 &  7.671e$-$2 &  5.226e$-$2 & 4.217e$-$3 &  1.420e$-$3 \\
$[$\oiii$]$ $\lambda$4363             &  2.603e$-$3 &  6.414e$-$2 & 1.776e$-$1 &  3.002e$-$1 &  1.303e$-$3 &  3.875e$-$2 & 1.161e$-$1 &  2.366e$-$1 \\
\heii\ $\lambda$4686                  &  2.920e$-$1 &  3.274e$-$1 & 3.043e$-$1 &  1.364e$-$1 &  2.275e$-$1 &  2.647e$-$1 & 2.524e$-$1 &  1.132e$-$1 \\
$[$\oiii$]$ $\lambda$4959             &  3.065e$+$0 &  4.525e$+$0 & 1.334e$+$0 &  4.306e$-$2 &  1.974e$+$0 &  2.813e$+$0 & 9.373e$-$1 &  3.617e$-$2 \\
$[$\oiii$]$ $\lambda$5007             &  8.826e$+$0 &  1.303e$+$1 & 3.841e$+$0 &  1.240e$-$1 &  5.685e$+$0 &  8.101e$+$0 & 2.699e$+$0 &  1.042e$-$1 \\
\end{tabular}
}
\end{table}

\clearpage

%
%
\begin{table}[!hp]
{\scriptsize
\caption{Dust-free IR line ratios wrt H$\beta$, $0.25 \times Z_\odot$}
\begin{tabular}{c|cccc|cccc}
\tableline
\tableline
& & $\alpha = -1.2$ & & & & $\alpha = -1.4$ & & \\
\tableline
 $\log U_0$ & -1.0 & -2.0 & -3.0 & -4.0 & -1.0 & -2.0 & -3.0 & -4.0 \\ 
Br$\alpha$ $\lambda$4.051    & 5.819e$-$2 & 6.462e$-$2 & 7.327e$-$2 & 8.064e$-$2 & 6.087e$-$2 & 6.571e$-$2 & 7.326e$-$2 & 8.003e$-$2  \\ 
$[$\arvi$]$ $\lambda$4.530   & 8.949e$-$3 & 9.005e$-$4 & 0.000e$+$0 & 0.000e$+$0 & 7.078e$-$3 & 5.064e$-$4 & 0.000e$+$0 & 0.000e$+$0  \\ 
$[$\arii$]$ $\lambda$6.983   & 1.132e$-$3 & 7.294e$-$3 & 2.367e$-$2 & 5.422e$-$2 & 8.185e$-$4 & 5.249e$-$3 & 1.734e$-$2 & 4.191e$-$2  \\ 
$[$\ariii$]$ $\lambda$8.983  & 3.142e$-$1 & 3.925e$-$2 & 3.902e$-$5 & 0.000e$+$0 & 2.294e$-$1 & 1.956e$-$2 & 1.740e$-$5 & 0.000e$+$0  \\ 
$[$\nevi$]$ $\lambda$7.652   & 7.458e$-$3 & 4.097e$-$2 & 6.394e$-$2 & 4.018e$-$2 & 7.966e$-$3 & 4.360e$-$2 & 6.661e$-$2 & 4.208e$-$2  \\ 
$[$\siv$]$ $\lambda$10.44    & 8.766e$-$1 & 7.646e$-$1 & 2.230e$-$1 & 3.026e$-$3 & 1.027e$+$0 & 8.820e$-$1 & 2.021e$-$1 & 2.595e$-$3  \\
\end{tabular}
}
\end{table}
%
%
\begin{table}[!hp]
{\scriptsize
\caption{Dust-free IR line ratios wrt H$\beta$, $0.5 \times Z_\odot$}
\begin{tabular}{c|cccc|cccc}
\tableline
\tableline \vspace{1mm}
& & $\alpha = -1.2$ & & & & $\alpha = -1.4$ & & \\
\tableline
 $\log U_0$ & -1.0 & -2.0 & -3.0 & -4.0 & -1.0 & -2.0 & -3.0 & -4.0 \\ 
Br$\alpha$ $\lambda$4.051    & 6.189e$-$2 & 6.664e$-$2 & 7.467e$-$2 & 8.267e$-$2 & 6.488e$-$2 & 6.817e$-$2 & 7.483e$-$2 & 8.173e$-$2  \\ 
$[$\arvi$]$ $\lambda$4.530   & 1.462e$-$2 & 1.562e$-$3 & 0.000e$+$0 & 0.000e$+$0 & 1.125e$-$2 & 8.569e$-$4 & 0.000e$+$0 & 0.000e$+$0  \\ 
$[$\arii$]$ $\lambda$6.983   & 1.881e$-$3 & 1.322e$-$2 & 4.247e$-$2 & 9.450e$-$2 & 1.336e$-$3 & 9.445e$-$3 & 3.088e$-$2 & 7.360e$-$2  \\ 
$[$\ariii$]$ $\lambda$8.983  & 5.531e$-$1 & 6.565e$-$2 & 6.782e$-$5 & 0.000e$+$0 & 3.858e$-$1 & 3.154e$-$2 & 2.954e$-$5 & 0.000e$+$0  \\ 
$[$\nevi$]$ $\lambda$7.652   & 1.468e$-$2 & 8.242e$-$2 & 1.262e$-$1 & 7.822e$-$2 & 1.576e$-$2 & 8.661e$-$2 & 1.306e$-$1 & 8.147e$-$2  \\ 
$[$\siv$]$ $\lambda$10.44    & 1.707e$+$0 & 1.526e$+$0 & 4.246e$-$1 & 5.873e$-$3 & 1.953e$+$0 & 1.713e$+$0 & 3.793e$-$1 & 4.999e$-$3  \\
\end{tabular}
}
\end{table}
%
%
\begin{table}[!hp]
{\scriptsize
\caption{Dust-free IR line ratios wrt H$\beta$, $1 \times Z_\odot$}
\begin{tabular}{c|cccc|cccc}
\tableline
\tableline \vspace{1mm}
& & $\alpha = -1.2$ & & & & $\alpha = -1.4$ & & \\
\tableline
 $\log U_0$ & -1.0 & -2.0 & -3.0 & -4.0 & -1.0 & -2.0 & -3.0 & -4.0 \\ 
Br$\alpha$ $\lambda$4.051    & 6.737e$-$2 & 7.021e$-$2 & 7.744e$-$2 & 8.597e$-$2 & 7.069e$-$2 & 7.250e$-$2 & 7.803e$-$2 & 8.472e$-$2  \\ 
$[$\arvi$]$ $\lambda$4.530   & 2.138e$-$2 & 2.400e$-$3 & 0.000e$+$0 & 0.000e$+$0 & 1.639e$-$2 & 1.270e$-$3 & 0.000e$+$0 & 0.000e$+$0  \\ 
$[$\arii$]$ $\lambda$6.983   & 2.583e$-$3 & 2.216e$-$2 & 7.179e$-$2 & 1.573e$-$1 & 1.799e$-$3 & 1.567e$-$2 & 5.189e$-$2 & 1.246e$-$1  \\ 
$[$\ariii$]$ $\lambda$8.983  & 8.689e$-$1 & 9.545e$-$2 & 1.017e$-$4 & 0.000e$+$0 & 5.785e$-$1 & 4.355e$-$2 & 4.222e$-$5 & 0.000e$+$0  \\ 
$[$\nevi$]$ $\lambda$7.652   & 2.971e$-$2 & 1.634e$-$1 & 2.428e$-$1 & 1.460e$-$1 & 3.204e$-$2 & 1.682e$-$1 & 2.475e$-$1 & 1.503e$-$1  \\ 
$[$\siv$]$ $\lambda$10.44    & 3.386e$+$0 & 2.972e$+$0 & 7.600e$-$1 & 1.088e$-$2 & 3.744e$+$0 & 3.187e$+$0 & 6.587e$-$1 & 9.109e$-$3  \\
\end{tabular}
}
\end{table}
%
%
\begin{table}[!hp]
{\scriptsize
\caption{Dust-free IR line ratios wrt H$\beta$, $2 \times Z_\odot$}
\begin{tabular}{c|cccc|cccc}
\tableline
\tableline \vspace{1mm}
& & $\alpha = -1.2$ & & & & $\alpha = -1.4$ & & \\
\tableline
 $\log U_0$ & -1.0 & -2.0 & -3.0 & -4.0 & -1.0 & -2.0 & -3.0 & -4.0 \\ 
Br$\alpha$ $\lambda$4.051    & 7.581e$-$2 & 7.631e$-$2 & 8.303e$-$2 & 9.206e$-$2 & 7.969e$-$2 & 7.960e$-$2 & 8.457e$-$2 & 9.094e$-$2  \\ 
$[$\arvi$]$ $\lambda$4.530   & 2.896e$-$2 & 3.119e$-$3 & 0.000e$+$0 & 0.000e$+$0 & 2.247e$-$2 & 1.589e$-$3 & 0.000e$+$0 & 0.000e$+$0  \\ 
$[$\arii$]$ $\lambda$6.983   & 2.794e$-$3 & 3.371e$-$2 & 1.141e$-$1 & 2.526e$-$1 & 1.892e$-$3 & 2.373e$-$2 & 8.279e$-$2 & 2.056e$-$1  \\ 
$[$\ariii$]$ $\lambda$8.983  & 1.193e$+$0 & 1.159e$-$1 & 1.185e$-$4 & 0.000e$+$0 & 7.592e$-$1 & 4.985e$-$2 & 4.616e$-$5 & 0.000e$+$0  \\ 
$[$\nevi$]$ $\lambda$7.652   & 6.266e$-$2 & 3.139e$-$1 & 4.385e$-$1 & 2.441e$-$1 & 6.813e$-$2 & 3.169e$-$1 & 4.373e$-$1 & 2.450e$-$1  \\ 
$[$\siv$]$ $\lambda$10.44    & 6.637e$+$0 & 5.472e$+$0 & 1.208e$+$0 & 1.800e$-$2 & 7.019e$+$0 & 5.565e$+$0 & 1.004e$+$0 & 1.459e$-$2  \\
\end{tabular}
}
\end{table}
%
%
\begin{table}[!hp]
{\scriptsize
\caption{Dust-free IR line ratios wrt H$\beta$, $4 \times Z_\odot$\label{tab:ndIR4}}
\begin{tabular}{c|cccc|cccc}
\tableline
\tableline \vspace{1mm}
& & $\alpha = -1.2$ & & & & $\alpha = -1.4$ & & \\
\tableline
 $\log U_0$ & -1.0 & -2.0 & -3.0 & -4.0 & -1.0 & -2.0 & -3.0 & -4.0 \\ 
Br$\alpha$ $\lambda$4.051    & 9.124e$-$2 & 8.575e$-$2 & 9.158e$-$2 & 1.015e$-$1 & 9.702e$-$2 & 9.058e$-$2 & 9.364e$-$2 & 9.985e$-$2  \\ 
$[$\arvi$]$ $\lambda$4.530   & 3.737e$-$2 & 3.348e$-$3 & 1.033e$-$6 & 0.000e$+$0 & 2.889e$-$2 & 1.642e$-$3 & 0.000e$+$0 & 0.000e$+$0  \\ 
$[$\arii$]$ $\lambda$6.983   & 2.546e$-$3 & 4.794e$-$2 & 1.788e$-$1 & 3.988e$-$1 & 1.839e$-$3 & 3.470e$-$2 & 1.339e$-$1 & 3.359e$-$1  \\ 
$[$\ariii$]$ $\lambda$8.983  & 1.433e$+$0 & 1.167e$-$1 & 1.060e$-$4 & 0.000e$+$0 & 8.734e$-$1 & 4.696e$-$2 & 3.896e$-$5 & 0.000e$+$0  \\ 
$[$\nevi$]$ $\lambda$7.652   & 1.422e$-$1 & 5.838e$-$1 & 7.211e$-$1 & 3.329e$-$1 & 1.539e$-$1 & 5.713e$-$1 & 7.007e$-$1 & 3.266e$-$1  \\ 
$[$\siv$]$ $\lambda$10.44    & 1.147e$+$1 & 9.238e$+$0 & 1.688e$+$0 & 2.554e$-$2 & 1.129e$+$1 & 8.757e$+$0 & 1.347e$+$0 & 2.011e$-$2  \\
\end{tabular}
}
\end{table}
\clearpage
%
%
\begin{table}[!hp]
{\scriptsize
\caption{Undepleted dust-free UV line line ratios wrt H$\beta$, $0.25 \times Z_\odot$\label{tab:ndunDUV0.25}}
\begin{tabular}{c|cccc|cccc}
\tableline
\tableline
& & & $\alpha = -1.2$ & & & & $\alpha = -1.4$ & \\
\tableline
$\log U_0$ & -1.0 & -2.0 & -3.0 & -4.0 & -1.0 & -2.0 & -3.0 & -4.0 \\  
\ciii\ $\lambda$977           &  3.433e+00 &  1.504e-01 & 4.512e-03 &  0.000e+00 &  2.047e+00 &  1.885e-01 & 4.881e-03 &  0.000e+00 \\ 
\niii\ $\lambda$991            &  2.998e-01 &  2.379e-02 & 3.692e-02 &  5.643e-05 &  2.401e-01 &  1.663e-02 & 3.569e-02 &  5.087e-05 \\ 
\ovi\ $\lambda\lambda$1032,7  &  2.051e-01 &  1.866e-01 & 1.693e-03 &  6.824e-01 &  1.763e-01 &  1.678e-01 & 4.383e-03 &  5.786e-01 \\ 
\nv\ $\lambda$1240            &  2.268e-01 &  8.762e-02 & 7.024e-02 &  9.026e-04 &  1.243e-01 &  1.083e-01 & 5.831e-02 &  1.005e-03 \\ 
\siliv+\oiv$]$ $\lambda$1400  &  9.418e+00 &  7.372e+00 & 2.662e+00 &  1.966e+00 &  9.512e+00 &  6.112e+00 & 2.411e+00 &  2.030e+00 \\ 
\niv$]$ $\lambda$1486         &  2.235e+00 &  3.517e-02 & 3.448e-01 &  3.751e-01 &  1.788e+00 &  3.821e-02 & 3.186e-01 &  3.408e-01 \\
\end{tabular}
}
\end{table}
%
%
\begin{table}[!hp]
{\scriptsize
\caption{Undepleted dust-free UV line line ratios wrt H$\beta$, $0.5 \times Z_\odot$}
\begin{tabular}{c|cccc|cccc}
\tableline
\tableline \vspace{1mm}
& & & $\alpha = -1.2$ & & & & $\alpha = -1.4$ & \\
\tableline
$\log U_0$ & -1.0 & -2.0 & -3.0 & -4.0 &  -1.0 & -2.0 & -3.0 & -4.0 \\
\ciii\ $\lambda$977           &  3.823e+00 &  1.774e-01 & 4.475e-03 &  0.000e+00 &  2.032e+00 &  1.684e-01 & 2.849e-03 &  0.000e+00 \\ 
\niii\ $\lambda$991            &  5.408e-01 &  2.601e-02 & 4.133e-02 &  6.258e-05 &  3.858e-01 &  1.632e-02 & 3.164e-02 &  4.574e-05 \\ 
\ovi\ $\lambda\lambda$1032,7  &  3.211e-01 &  2.036e-01 & 2.442e-03 &  7.043e-01 &  2.364e-01 &  1.276e-01 & 4.204e-03 &  5.973e-01 \\ 
\nv\ $\lambda$1240            &  2.121e-01 &  1.028e-01 & 7.592e-02 &  1.062e-03 &  1.256e-01 &  9.543e-02 & 5.003e-02 &  9.748e-04 \\ 
\siliv+\oiv$]$ $\lambda$1400  &  9.325e+00 &  8.380e+00 & 2.980e+00 &  2.528e+00 &  8.050e+00 &  5.926e+00 & 2.505e+00 &  2.321e+00 \\ 
\niv$]$ $\lambda$1486         &  2.240e+00 &  6.230e-02 & 5.212e-01 &  5.903e-01 &  1.755e+00 &  4.736e-02 & 4.251e-01 &  4.989e-01 \\
\end{tabular}
}
\end{table}
%
%
\begin{table}[!hp]
{\scriptsize
\caption{Undepleted dust-free UV line line ratios wrt H$\beta$, $1 \times Z_\odot$}
\begin{tabular}{c|cccc|cccc}
\tableline
\tableline \vspace{1mm}
& & & $\alpha = -1.2$ & & & & $\alpha = -1.4$ & \\
\tableline
$\log U_0$ & -1.0 & -2.0 & -3.0 & -4.0 & -1.0 & -2.0 & -3.0 & -4.0 \\ 
\ciii\ $\lambda$977           &  3.528e+00 &  1.221e-01 & 2.031e-03 &  0.000e+00 &  1.702e+00 &  9.078e-02 & 9.884e-04 &  0.000e+00 \\ 
\niii\ $\lambda$991            &  9.113e-01 &  2.027e-02 & 2.143e-02 &  3.048e-05 &  5.961e-01 &  1.067e-02 & 1.217e-02 &  1.851e-05 \\ 
\ovi\ $\lambda\lambda$1032,7  &  4.858e-01 &  1.584e-01 & 3.730e-03 &  7.331e-01 &  3.211e-01 &  8.318e-02 & 1.024e-03 &  6.146e-01 \\ 
\nv\ $\lambda$1240            &  1.621e-01 &  7.152e-02 & 3.995e-02 &  5.867e-04 &  9.382e-02 &  1.078e-01 & 2.035e-02 &  4.611e-04 \\ 
\siliv+\oiv$]$ $\lambda$1400  &  7.771e+00 &  7.330e+00 & 2.665e+00 &  2.335e+00 &  6.230e+00 &  4.591e+00 & 2.079e+00 &  2.026e+00 \\ 
\niv$]$ $\lambda$1486         &  2.188e+00 &  8.144e-02 & 5.931e-01 &  6.834e-01 &  1.694e+00 &  5.035e-02 & 4.215e-01 &  5.363e-01 \\
\end{tabular}
}
\end{table}
%
%
\begin{table}[!hp]
{\scriptsize
\caption{Undepleted dust-free UV line line ratios wrt H$\beta$, $2 \times Z_\odot$}
\begin{tabular}{c|cccc|cccc}
\tableline
\tableline \vspace{1mm}
& & & $\alpha = -1.2$ & & & & $\alpha = -1.4$ & \\
\tableline
$\log U_0$ & -1.0 & -2.0 & -3.0 & -4.0 & -1.0 & -2.0 & -3.0 & -4.0 \\ 
\ciii\ $\lambda$977           &  2.818e+00 &  5.373e-02 & 2.965e-04 &  0.000e+00 &  1.231e+00 &  3.410e-02 & 1.051e-04 &  0.000e+00 \\ 
\niii\ $\lambda$991            &  1.490e+00 &  1.159e-02 & 3.169e-03 &  3.561e-06 &  9.007e-01 &  4.767e-03 & 1.231e-03 &  1.671e-06 \\ 
\ovi\ $\lambda\lambda$1032,7  &  7.632e-01 &  9.059e-02 & 8.525e-04 &  7.538e-01 &  4.631e-01 &  3.718e-02 & 3.250e-04 &  6.243e-01 \\ 
\nv\ $\lambda$1240            &  1.132e-01 &  9.370e-02 & 7.117e-03 &  8.916e-05 &  6.670e-02 &  4.487e-02 & 2.655e-03 &  5.640e-05 \\ 
\siliv+\oiv$]$ $\lambda$1400  &  6.644e+00 &  5.305e+00 & 2.032e+00 &  1.035e+00 &  5.027e+00 &  3.092e+00 & 1.595e+00 &  8.011e-01 \\ 
\niv$]$ $\lambda$1486         &  2.112e+00 &  8.131e-02 & 4.008e-01 &  3.901e-01 &  1.620e+00 &  4.027e-02 & 2.155e-01 &  2.679e-01 \\
\end{tabular}
}
\end{table}
%
%
\begin{table}[!hp]
{\scriptsize
\caption{Undepleted dust-free UV line line ratios wrt H$\beta$, $4 \times Z_\odot$}
\begin{tabular}{c|cccc|cccc}
\tableline
\tableline \vspace{1mm}
& & & $\alpha = -1.2$ & & & & $\alpha = -1.4$ & \\
\tableline
$\log U_0$ & -1.0 & -2.0 & -3.0 & -4.0 & -1.0 & -2.0 & -3.0 & -4.0 \\ 
\ciii\ $\lambda$977           &  2.025e+00 &  2.625e-02 & 1.425e-05 &  0.000e+00 &  7.939e-01 &  1.223e-02 & 3.129e-06 &  0.000e+00 \\ 
\niii\ $\lambda$991            &  2.369e+00 &  5.386e-03 & 1.344e-04 &  0.000e+00 &  1.319e+00 &  1.495e-03 & 3.028e-05 &  0.000e+00 \\ 
\ovi\ $\lambda\lambda$1032,7  &  1.210e+00 &  4.235e-02 & 7.978e-05 &  7.467e-01 &  6.762e-01 &  1.172e-02 & 1.958e-05 &  6.090e-01 \\ 
\nv\ $\lambda$1240            &  8.232e-02 &  1.575e-02 & 4.294e-04 &  1.760e-06 &  4.971e-02 &  1.492e-02 & 1.009e-04 &  0.000e+00 \\ 
\siliv+\oiv$]$ $\lambda$1400  &  6.192e+00 &  3.801e+00 & 1.841e+00 &  1.473e-01 &  4.311e+00 &  2.221e+00 & 1.488e+00 &  9.688e-02 \\ 
\niv$]$ $\lambda$1486         &  2.049e+00 &  7.734e-02 & 1.097e-01 &  8.298e-02 &  1.554e+00 &  2.819e-02 & 3.151e-02 &  4.996e-02 \\
\end{tabular}
}
\end{table}

\clearpage
%
%
\begin{table}[!hp]
{\scriptsize
\caption{Undepleted dust-free Visible line ratios wrt H$\beta$, $0.25 \times Z_\odot$}
\begin{tabular}{c|cccc|cccc}
\tableline
\tableline
& & & $\alpha = -1.2$ & & & & $\alpha = -1.4$ & \\
\tableline
$\log U_0$ & -1.0 & -2.0 & -3.0 & -4.0 & -1.0 & -2.0 & -3.0 & -4.0 \\
$[$\oii$]$ $\lambda\lambda$3727,9     &  7.521e-02 &  5.684e-01 & 2.527e+00 &  2.548e+00 &  3.596e-02 &  2.730e-01 & 1.273e+00 &  1.408e+00 \\ 
$[$\sii$]$ $\lambda\lambda$4067,76    &  3.402e-01 &  2.002e-01 & 4.613e-02 &  4.606e-04 &  2.983e-01 &  2.015e-01 & 4.966e-02 &  2.736e-04 \\ 
$[$\oiii$]$ $\lambda$4363             &  2.502e-03 &  1.265e-02 & 4.475e-02 &  1.039e-01 &  3.614e-03 &  1.878e-02 & 7.429e-02 &  1.980e-01 \\ 
\heii\ $\lambda$4686                  &  2.804e-01 &  2.905e-01 & 2.358e-01 &  1.050e-01 &  2.264e-01 &  2.350e-01 & 1.970e-01 &  8.832e-02 \\ 
$[$\oiii$]$ $\lambda$4959             &  3.403e+00 &  2.759e+00 & 1.105e+00 &  3.367e-02 &  3.424e+00 &  2.896e+00 & 1.157e+00 &  3.574e-02 \\ 
$[$\oiii$]$ $\lambda$5007             &  9.802e+00 &  7.946e+00 & 3.181e+00 &  9.698e-02 &  9.861e+00 &  8.341e+00 & 3.333e+00 &  1.029e-01 \\
\end{tabular}
}
\end{table}
%
%
\begin{table}[!hp]
{\scriptsize
\caption{Undepleted dust-free Visible line ratios wrt H$\beta$, $0.5 \times Z_\odot$}
\begin{tabular}{c|cccc|cccc}
\tableline
\tableline \vspace{1mm}
& & & $\alpha = -1.2$ & & & & $\alpha = -1.4$ & \\
\tableline
$\log U_0$ & -1.0 & -2.0 & -3.0 & -4.0 & -1.0 & -2.0 & -3.0 & -4.0 \\ 
$[$\oii$]$ $\lambda\lambda$3727,9     &  1.173e-01 &  9.848e-01 & 4.253e+00 &  4.123e+00 &  1.050e-01 &  8.548e-01 & 4.029e+00 &  4.088e+00 \\ 
$[$\sii$]$ $\lambda\lambda$4067,76    &  3.470e-01 &  2.595e-01 & 6.220e-02 &  8.611e-04 &  2.517e-01 &  2.083e-01 & 5.510e-02 &  8.569e-04 \\ 
$[$\oiii$]$ $\lambda$4363             &  3.810e-03 &  2.065e-02 & 6.903e-02 &  1.583e-01 &  2.935e-03 &  1.643e-02 & 5.992e-02 &  1.519e-01 \\ 
\heii\ $\lambda$4686                  &  2.874e-01 &  2.969e-01 & 2.427e-01 &  1.092e-01 &  2.288e-01 &  2.410e-01 & 2.045e-01 &  9.278e-02 \\ 
$[$\oiii$]$ $\lambda$4959             &  4.797e+00 &  4.477e+00 & 1.807e+00 &  5.498e-02 &  4.389e+00 &  4.221e+00 & 1.748e+00 &  5.468e-02 \\ 
$[$\oiii$]$ $\lambda$5007             &  1.382e+01 &  1.289e+01 & 5.204e+00 &  1.583e-01 &  1.264e+01 &  1.216e+01 & 5.034e+00 &  1.575e-01 \\
\end{tabular}
}
\end{table}
%
%
\begin{table}[!hp]
{\scriptsize
\caption{Undepleted dust-free Visible line ratios wrt H$\beta$, $1 \times Z_\odot$}
\begin{tabular}{c|cccc|cccc}
\tableline
\tableline \vspace{1mm}
& & & $\alpha = -1.2$ & & & & $\alpha = -1.4$ & \\
\tableline
$\log U_0$ & -1.0 & -2.0 & -3.0 & -4.0 & -1.0 & -2.0 & -3.0 & -4.0 \\ 
$[$\oii$]$ $\lambda\lambda$3727,9     &  1.449e-01 &  1.411e+00 & 5.630e+00 &  4.889e+00 &  1.287e-01 &  1.117e+00 & 4.816e+00 &  4.493e+00 \\ 
$[$\sii$]$ $\lambda\lambda$4067,76    &  2.599e-01 &  2.325e-01 & 5.157e-02 &  1.335e-03 &  1.656e-01 &  1.537e-01 & 3.806e-02 &  1.269e-03 \\ 
$[$\oiii$]$ $\lambda$4363             &  4.197e-03 &  2.954e-02 & 9.301e-02 &  1.972e-01 &  2.996e-03 &  2.206e-02 & 7.780e-02 &  1.814e-01 \\ 
\heii\ $\lambda$4686                  &  2.875e-01 &  3.029e-01 & 2.553e-01 &  1.157e-01 &  2.265e-01 &  2.454e-01 & 2.150e-01 &  9.745e-02 \\ 
$[$\oiii$]$ $\lambda$4959             &  5.504e+00 &  5.909e+00 & 2.297e+00 &  6.707e-02 &  4.620e+00 &  4.928e+00 & 2.012e+00 &  6.253e-02 \\ 
$[$\oiii$]$ $\lambda$5007             &  1.585e+01 &  1.702e+01 & 6.614e+00 &  1.932e-01 &  1.331e+01 &  1.419e+01 & 5.794e+00 &  1.801e-01 \\
\end{tabular}
}
\end{table}
%
%
\begin{table}[!hp]
{\scriptsize
\caption{Undepleted dust-free Visible line ratios wrt H$\beta$, $2 \times Z_\odot$}
\begin{tabular}{c|cccc|cccc}
\tableline
\tableline \vspace{1mm}
& & & $\alpha = -1.2$ & & & & $\alpha = -1.4$ & \\
\tableline
$\log U_0$ & -1.0 & -2.0 & -3.0 & -4.0 & -1.0 & -2.0 & -3.0 & -4.0 \\ 
$[$\oii$]$ $\lambda\lambda$3727,9     &  1.025e-01 &  1.353e+00 & 4.569e+00 &  2.974e+00 &  8.820e-02 &  8.958e-01 & 3.295e+00 &  2.420e+00 \\ 
$[$\sii$]$ $\lambda\lambda$4067,76    &  1.496e-01 &  1.405e-01 & 1.980e-02 &  1.411e-03 &  8.313e-02 &  7.532e-02 & 1.171e-02 &  1.360e-03 \\ 
$[$\oiii$]$ $\lambda$4363             &  2.071e-03 &  3.280e-02 & 9.314e-02 &  1.397e-01 &  1.555e-03 &  2.114e-02 & 6.867e-02 &  1.150e-01 \\ 
\heii\ $\lambda$4686                  &  2.840e-01 &  3.104e-01 & 2.752e-01 &  1.234e-01 &  2.217e-01 &  2.521e-01 & 2.305e-01 &  1.029e-01 \\ 
$[$\oiii$]$ $\lambda$4959             &  4.555e+00 &  5.934e+00 & 1.953e+00 &  5.022e-02 &  3.381e+00 &  4.278e+00 & 1.520e+00 &  4.310e-02 \\ 
$[$\oiii$]$ $\lambda$5007             &  1.312e+01 &  1.709e+01 & 5.624e+00 &  1.446e-01 &  9.738e+00 &  1.232e+01 & 4.377e+00 &  1.241e-01 \\
\end{tabular}
}
\end{table}
%
%
\begin{table}[!hp]
{\scriptsize
\caption{Undepleted dust-free Visible line ratios wrt H$\beta$, $4 \times Z_\odot$}
\begin{tabular}{c|cccc|cccc}
\tableline
\tableline \vspace{1mm}
& & & $\alpha = -1.2$ & & & & $\alpha = -1.4$ & \\
\tableline
$\log U_0$ & -1.0 & -2.0 & -3.0 & -4.0 & -1.0 & -2.0 & -3.0 & -4.0 \\ 
$[$\oii$]$ $\lambda\lambda$3727,9     &  1.949e-02 &  6.496e-01 & 1.818e+00 &  7.229e-01 &  1.141e-02 &  2.962e-01 & 9.829e-01 &  5.133e-01 \\ 
$[$\sii$]$ $\lambda\lambda$4067,76    &  9.184e-02 &  7.188e-02 & 3.350e-03 &  7.038e-04 &  4.933e-02 &  2.986e-02 & 1.419e-03 &  2.618e-04 \\ 
$[$\oiii$]$ $\lambda$4363             &  2.530e-04 &  2.207e-02 & 5.401e-02 &  4.302e-02 &  1.059e-04 &  1.013e-02 & 2.941e-02 &  3.132e-02 \\ 
\heii\ $\lambda$4686                  &  2.812e-01 &  3.234e-01 & 2.980e-01 &  1.281e-01 &  2.171e-01 &  2.642e-01 & 2.462e-01 &  1.053e-01 \\ 
$[$\oiii$]$ $\lambda$4959             &  2.539e+00 &  4.409e+00 & 1.055e+00 &  2.018e-02 &  1.634e+00 &  2.504e+00 & 6.859e-01 &  1.542e-02 \\ 
$[$\oiii$]$ $\lambda$5007             &  7.312e+00 &  1.270e+01 & 3.039e+00 &  5.811e-02 &  4.706e+00 &  7.212e+00 & 1.976e+00 &  4.442e-02 \\
\end{tabular}
}
\end{table}

\clearpage
%
%
\begin{table}[!hp]
{\scriptsize
\caption{Undepleted dust-free IR line ratios wrt H$\beta$, $0.25 \times Z_\odot$}
\begin{tabular}{c|cccc|cccc}
\tableline
\tableline
& & & $\alpha = -1.2$ & & & & $\alpha = -1.4$ & \\
\tableline
 $\log U_0$ & -1.0 & -2.0 & -3.0 & -4.0 & -1.0 & -2.0 & -3.0 & -4.0 \\ 
Br$\alpha$ $\lambda$4.051    & 6.039e-02 & 6.596e-02 & 7.452e-02 & 8.328e-02  & 6.238e-02 & 6.610e-02 & 7.275e-02 & 7.951e-02  \\ 
$[$\arvi$]$ $\lambda$4.530   & 8.136e-03 & 8.660e-04 & 0.000e+00 & 0.000e+00  & 6.341e-03 & 4.854e-04 & 0.000e+00 & 0.000e+00  \\ 
$[$\arii$]$ $\lambda$6.983   & 1.043e-03 & 6.903e-03 & 2.136e-02 & 4.531e-02  & 7.575e-04 & 5.059e-03 & 1.641e-02 & 3.876e-02  \\ 
$[$\ariii$]$ $\lambda$8.983  & 3.066e-01 & 3.678e-02 & 3.714e-05 & 0.000e+00  & 2.169e-01 & 1.813e-02 & 1.662e-05 & 0.000e+00  \\ 
$[$\nevi$]$ $\lambda$7.652   & 7.238e-03 & 4.091e-02 & 6.353e-02 & 4.014e-02  & 7.261e-03 & 4.029e-02 & 6.146e-02 & 3.920e-02  \\ 
$[$\siv$]$ $\lambda$10.44    & 8.303e-01 & 7.548e-01 & 2.187e-01 & 2.993e-03  & 7.110e-01 & 6.289e-01 & 1.412e-01 & 1.860e-03  \\
\end{tabular}
}
\end{table}
%
%
\begin{table}[!hp]
{\scriptsize
\caption{Undepleted dust-free IR line ratios wrt H$\beta$, $0.5 \times Z_\odot$}
\begin{tabular}{c|cccc|cccc}
\tableline
\tableline \vspace{1mm}
& & & $\alpha = -1.2$ & & & & $\alpha = -1.4$ & \\
\tableline
 $\log U_0$ & -1.0 & -2.0 & -3.0 & -4.0 & -1.0 & -2.0 & -3.0 & -4.0 \\ 
Br$\alpha$ $\lambda$4.051    & 6.483e-02 & 6.901e-02 & 7.752e-02 & 8.786e-02  & 6.798e-02 & 7.081e-02 & 7.765e-02 & 8.531e-02  \\ 
$[$\arvi$]$ $\lambda$4.530   & 1.266e-02 & 1.450e-03 & 0.000e+00 & 0.000e+00  & 9.752e-03 & 7.787e-04 & 0.000e+00 & 0.000e+00  \\ 
$[$\arii$]$ $\lambda$6.983   & 1.529e-03 & 1.186e-02 & 3.586e-02 & 7.306e-02  & 1.043e-03 & 8.261e-03 & 2.550e-02 & 5.828e-02  \\ 
$[$\ariii$]$ $\lambda$8.983  & 5.220e-01 & 5.836e-02 & 6.113e-05 & 0.000e+00  & 3.526e-01 & 2.713e-02 & 2.611e-05 & 0.000e+00  \\ 
$[$\nevi$]$ $\lambda$7.652   & 1.439e-02 & 8.143e-02 & 1.236e-01 & 7.663e-02  & 1.556e-02 & 8.475e-02 & 1.271e-01 & 7.941e-02  \\ 
$[$\siv$]$ $\lambda$10.44    & 1.622e+00 & 1.482e+00 & 4.054e-01 & 5.621e-03  & 1.852e+00 & 1.637e+00 & 3.579e-01 & 4.735e-03  \\
\end{tabular}
}
\end{table}
%
%
\begin{table}[!hp]
{\scriptsize
\caption{Undepleted dust-free IR line ratios wrt H$\beta$, $1 \times Z_\odot$}
\begin{tabular}{c|cccc|cccc}
\tableline
\tableline \vspace{1mm}
& & & $\alpha = -1.2$ & & & & $\alpha = -1.4$ & \\
\tableline
 $\log U_0$ & -1.0 & -2.0 & -3.0 & -4.0 & -1.0 & -2.0 & -3.0 & -4.0 \\ 
Br$\alpha$ $\lambda$4.051    & 7.129e-02 & 7.362e-02 & 8.232e-02 & 9.470e-02  & 7.463e-02 & 7.620e-02 & 8.282e-02 & 9.203e-02  \\ 
$[$\arvi$]$ $\lambda$4.530   & 1.835e-02 & 2.130e-03 & 0.000e+00 & 0.000e+00  & 1.439e-02 & 1.103e-03 & 0.000e+00 & 0.000e+00  \\ 
$[$\arii$]$ $\lambda$6.983   & 1.669e-03 & 1.828e-02 & 5.589e-02 & 1.129e-01  & 1.061e-03 & 1.251e-02 & 3.950e-02 & 9.284e-02  \\ 
$[$\ariii$]$ $\lambda$8.983  & 7.890e-01 & 7.953e-02 & 8.375e-05 & 0.000e+00  & 5.091e-01 & 3.497e-02 & 3.394e-05 & 0.000e+00  \\ 
$[$\nevi$]$ $\lambda$7.652   & 2.949e-02 & 1.587e-01 & 2.312e-01 & 1.363e-01  & 3.211e-02 & 1.622e-01 & 2.338e-01 & 1.386e-01  \\ 
$[$\siv$]$ $\lambda$10.44    & 3.240e+00 & 2.823e+00 & 6.924e-01 & 9.919e-03  & 3.566e+00 & 2.980e+00 & 5.913e-01 & 8.245e-03  \\
\end{tabular}
}
\end{table}
%
%
\begin{table}[!hp]
{\scriptsize
\caption{Undepleted dust-free IR line ratios wrt H$\beta$, $2 \times Z_\odot$}
\begin{tabular}{c|cccc|cccc}
\tableline
\tableline \vspace{1mm}
& & & $\alpha = -1.2$ & & & & $\alpha = -1.4$ & \\
\tableline
 $\log U_0$ & -1.0 & -2.0 & -3.0 & -4.0 & -1.0 & -2.0 & -3.0 & -4.0 \\ 
Br$\alpha$ $\lambda$4.051    & 8.081e-02 & 8.062e-02 & 9.000e-02 & 1.058e-01  & 8.461e-02 & 8.405e-02 & 9.097e-02 & 1.025e-01  \\ 
$[$\arvi$]$ $\lambda$4.530   & 2.560e-02 & 2.663e-03 & 0.000e+00 & 0.000e+00  & 2.017e-02 & 1.327e-03 & 0.000e+00 & 0.000e+00  \\ 
$[$\arii$]$ $\lambda$6.983   & 1.123e-03 & 2.504e-02 & 8.220e-02 & 1.687e-01  & 5.936e-04 & 1.695e-02 & 5.865e-02 & 1.440e-01  \\ 
$[$\ariii$]$ $\lambda$8.983  & 1.044e+00 & 9.110e-02 & 9.031e-05 & 0.000e+00  & 6.447e-01 & 3.758e-02 & 3.471e-05 & 0.000e+00  \\ 
$[$\nevi$]$ $\lambda$7.652   & 6.437e-02 & 3.023e-01 & 4.051e-01 & 2.118e-01  & 7.100e-02 & 3.038e-01 & 4.021e-01 & 2.113e-01  \\ 
$[$\siv$]$ $\lambda$10.44    & 6.315e+00 & 5.112e+00 & 1.057e+00 & 1.825e-02  & 6.638e+00 & 5.125e+00 & 8.719e-01 & 1.581e-02  \\
\end{tabular}
}
\end{table}
%
%
\begin{table}[!hp]
{\scriptsize
\caption{Undepleted dust-free IR line ratios wrt H$\beta$, $4 \times Z_\odot$\label{tab:ndunDIR4}}
%
}
\end{table}

\clearpage
%
%
\begin{figure}[!ht]
\caption{\label{N/Ofig}[N/O] ratio against Oxygen abundance for a sample
of galaxies and \hii\ regions, and our analytic fit to the [N/O]
variation with metallicity. The Solar value is marked by the square
and constrains our analytic fit. The data sets are as marked in the
key, with the references given in the text.}
\end{figure}

\begin{figure}[!ht]
\caption{\label{fig:ext} The extinction and scattering cross-section
of dust per H atom in a 1$Z_\odot$ metallicity model. Note the peak in
extinction curve around the Lyman limit (13.6 eV).}
\end{figure} 

\begin{figure}[!ht]
\caption{\label{fig:a-1.4Z1} Structure diagram showing the variation
of  electron  temperature,  electron  and  hydrogen  density  and  the
ionization states of hydrogen, helium, carbon and oxygen with distance
in a dusty  model. The distance is normalized to  1, with the distance
in cm given within the  figure. The model parameters are $1 Z_{\odot}$
Metallicity, $\alpha =-2.0$, n$_\mathrm{H}  = 1000$ \pccm, $\log U_0 =
-2.0$}
\end{figure}

\begin{figure}[!ht]
\caption{\label{fig:U-1} Structure diagram: $1 Z_{\odot}$ Metallicity,
$\alpha =-1.4$, n$_\mathrm{H} = 1000$ \pccm, $\log U_0= -1.0$} 
\end{figure}

\begin{figure}[!ht]
\caption{\label{fig:U-3} Structure diagram: $1 Z_{\odot}$ Metallicity,
$\alpha =-1.4$, n$_\mathrm{H} = 1000$ \pccm, $\log U_0= -3.0$} 
\end{figure}

\begin{figure}[!ht]
\caption{\label{fig:ndU-1} Structure  diagram for dust-free  model: $1
Z_{\odot}$ Metallicity,  $\alpha =-1.4$, n$_\mathrm{H}  = 1000$ \pccm,
$\log U_0= -1.0$} 
\end{figure}

\begin{figure}[!ht]
\caption{\label{fig:ndU-3} Structure  diagram for dust-free  model: $1
Z_{\odot}$ Metallicity,  $\alpha =-1.4$, n$_\mathrm{H}  = 1000$ \pccm,
$\log U_0= -3.0$} 
\end{figure}

\begin{figure}[!ht]
\caption{\label{fig:a-1.4Z0.25}  Structure  diagram: $0.25  Z_{\odot}$
Metallicity, $\alpha  =-1.4$, n$_\mathrm{H} = 1000$  \pccm, $\log U_0=
-2.0$} 
\end{figure}

\begin{figure}[!ht]
\caption{\label{fig:a-1.4Z4}   Structure    diagram:   $4   Z_{\odot}$
Metallicity, $\alpha  =-1.4$, n$_\mathrm{H} = 1000$  \pccm, $\log U_0=
-2.0$} 
\end{figure}

\begin{figure}[!ht]
\caption{\label{fig:a-1.2Z1}   Structure    diagram:   $1   Z_{\odot}$
Metallicity, $\alpha  =-1.2$, n$_\mathrm{H} = 1000$  \pccm, $\log U_0=
-2.0$} 
\end{figure}

\begin{figure}[!ht]
\caption{\label{fig:a-2.0Z1}   Structure    diagram:   $1   Z_{\odot}$
Metallicity, $\alpha  =-2.0$, n$_\mathrm{H} = 1000$  \pccm, $\log U_0=
-2.0$} 
\end{figure}

\begin{figure}[!ht]
\caption{\label{fig:dUl-2.0} $U_\mathrm{local}$ structure diagram showing the variation
of electron temperature, electron and hydrogen density and the
ionization states of hydrogen, helium, carbon and oxygen with local
ionization parameter in a dusty model. The model parameters are $1
Z_{\odot}$ Metallicity, $\alpha =-1.4$, n$_\mathrm{H} = 1000$ \pccm,
$\log U_0= -2.0$} 
\end{figure}

\begin{figure}[!ht]
\caption{\label{fig:ndUl-2.0} $U_\mathrm{local}$ structure diagram 
showing the dust-free equivalent of figure \ref{fig:dUl-2.0}, with the
same model parameters: $1
Z_{\odot}$ Metallicity, $\alpha =-1.4$, n$_\mathrm{H} = 1000$ \pccm,
$\log U_0= -2.0$} 
\end{figure}

\begin{figure}[!ht]
\caption{\label{fig:dUl-1.0} $U_\mathrm{local}$ structure diagram 
showing a dusty model with initial ionization parameter $\log U_0 =
-1.0$, and with  $1
Z_{\odot}$ Metallicity, $\alpha =-1.4$ and  n$_\mathrm{H} = 1000$ \pccm.} 
\end{figure}

\begin{figure}[!ht]
\caption{\label{fig:ndUl-1.0} $U_\mathrm{local}$ structure diagram 
showing a dust-free model with initial ionization parameter $\log U_0 =
-1.0$, and $1
Z_{\odot}$ Metallicity, $\alpha =-1.4$ and  n$_\mathrm{H} = 1000$ \pccm.} 
\end{figure}

\begin{figure}[!ht]
\caption{\label{fig:dUl0.0} $U_\mathrm{local}$ structure diagram 
showing a dusty model with initial ionization parameter $\log U_0 =
0.0$, and with  $1
Z_{\odot}$ Metallicity, $\alpha =-1.4$ and  n$_\mathrm{H} = 1000$ \pccm.} 
\end{figure}

\begin{figure}[!ht]
\caption{\label{fig:ndUl0.0} $U_\mathrm{local}$ structure diagram 
showing a dust-free model with initial ionization parameter $\log U_0 =
0.0$, and $1
Z_{\odot}$ Metallicity, $\alpha =-1.4$ and  n$_\mathrm{H} = 1000$ \pccm.} 
\end{figure}

\begin{figure}[!ht]
\caption{\label{fig:locspec} Variation of the ionizing spectrum ($\nu$
F$_{\nu}$) with $U_\mathrm{local}$ for both the dusty and dust-free
models. Curves are at $\log U_\mathrm{local} = -1.0, -2.0, -3.0$ and
$-4.0$, with decreasing flux in the 10 -- 100 eV range for each curve
respectively. Both models have $\log U_0 = 
-1.0$, $1Z_{\odot}$ Metallicity, $\alpha =-1.4$ and  n$_\mathrm{H} =
1000$ \pccm.}  
\end{figure}


\small
\begin{thebibliography}{}

\bibitem[Alexander et al.(2003)]{Alex03} Alexander, R.~D.,
Casali, M.~M., Andr{\' e}, P., Persi, P., \& Eiroa, C.\ 2003, \aap, 401,
613

\bibitem[Alexander et al.(2000)]{Alex00} Alexander, T., Lutz, D., Sturm, E., Genzel, R., Sternberg, A., \& Netzer, H.\ 2000, \apj, 536, 710 

\bibitem[Allen, Dopita, \& Tsvetanov(1998)]{ADT98} Allen, 
M.~G., Dopita, M.~A., \& Tsvetanov, Z.~I.\ 1998, \apj, 493, 571 (ADT)

\bibitem[Allen et al.(1999)]{Allen99}  Allen, M.~G., Dopita, M.~A.,
Tsvetanov, Z.~I. \& Sutherland, R.~S. 1999, \apj, 511, 686

\bibitem[Allende Prieto, Lambert, \& Asplund(2001)]{AllendeLamb01}
Allende Prieto, C., Lambert, D.~L., \& Asplund, M.\ 2001, \apjl, 556,
L63

\bibitem[Allende Prieto, Lambert, \& Asplund(2002)]{AllendeLamb02}
Allende Prieto, C., Lambert, D.~L., \& Asplund, M.\ 2002, \apjl, 573, L137

\bibitem[Anders \& Grevesse(1989)]{Anders89} Geochim. Cosmochim. Acta, 53,
197

\bibitem[Asplund(2000)]{Asplund00} Asplund, M.\ 2000, \aap, 359,
755

\bibitem[Asplund(2003)]{Asplund03} Asplund, M.\ 2003, ASP Conf.~Ser.:
CNO in the Universe, astro-ph/0302409 

\bibitem[Asplund, Nordlund, Trampedach, \& Stein(2000)]{AsplundNord00}
Asplund, M., Nordlund, {\AA}., Trampedach, R., \& Stein, R.~F.\ 2000,
\aap, 359, 743

\bibitem[Binette et~al.(1997)]{Ami97}  Binette, L., Wilson, A.~S., Raga, A.
\& Storchi-Bergmann, T., 1997, \aap, 327, 909

\bibitem[Binette et~al.(1996)]{Ami96}  Binette, L., Wilson, A.~S., \&
Storchi-Bergmann, T., 1996, \aap, 312, 365

\bibitem[Borkowski \& Harrington(1991)]{Borkowski91} Borkowski,
K.~J.~\& Harrington, J.~P.\ 1991, \apj, 379, 168

\bibitem[Bower et al.(1995)]{Bower95} Bower, G., Wilson, A., 
Morse, J.~A., Gelderman, R., Whittle, M., \& Mulchaey, J.\ 1995, \apj, 454, 
106 

\bibitem[Bradley, Kaiser \& Baan(2003)]{Bradley03} Bradley, L.~D., Kaiser, M.~E., \& Baan, W.~A., 2003, \apj accepted (astroph/0312067)

\bibitem[Brinkman  et~al.(2002)]{Brinkman02}Brinkman, A.~C., Kaastra, J.~S.,
vanderMeer, R.~L.~J., Kinkhabwala, A., Behar, E., Kahn, S.~M.,
Paerels, F.~B.~S., \& Sako, M. 2002, \aa, 396, 761

\bibitem[Cecil et al.(2002)]{Cecil02} Cecil, G., Dopita, M.~A., Groves, B., Wilson, A.~S., Ferruit, P., P{\' e}contal, E., \& Binette, L.\ 2002, \apj, 568, 627

\bibitem[Crenshaw \& Kraemer(2000)]{Crenshaw00} Crenshaw, D.~M.~\& Kraemer, S.~B.\ 2000, \apjl, 532, L101

\bibitem[Dopita, Binette \& Schwartz(1982)]{Dopita82}
Dopita, M.~A., Binette, L., \&  Schwartz, R.~D., 1982,
\apj, 261, 183

\bibitem[Dopita et al.(2002)]{DG02} Dopita, M.~A., Groves,
B.~A., Sutherland, R.~S., Binette, L., \& Cecil, G.\ 2002, \apj, 572,
753 (DG02)

\bibitem[Dopita et al.(2000)]{DopitaHII}
Dopita, M.~A., Kewley, L.~J., Heisler, C.~A., \& Sutherland, R.~S.\ 2000,
\apj, 542, 224

\bibitem[Dopita \& Sutherland(1995)]{DoSu95}  Dopita, M. A., \&
Sutherland,
R. S., 1995, \apj, 455, 468

\bibitem[Dopita \& Sutherland(1996)]{DoSu96}  Dopita, M. A., \&
Sutherland,
R. S., 1996, \apjs, 102, 161

\bibitem[Dopita \& Sutherland(2000)]{DopitaPE} Dopita, M.~A.~\&
Sutherland, R.~S.\ 2000, \apj, 539, 742

\bibitem[Draine(1978)]{Draine78} Draine, B.~T.\ 1978, \apjs, 36,
595

\bibitem[Draine(2003)]{Draine03} Draine, B.~T.\ 2003, \araa, 41, 
241

\bibitem[Draine \& Lazarian(1998)]{DraineLaz98} Draine, B.~T.~\& 
Lazarian, A.\ 1998, \apj, 508, 157

\bibitem[Draine \& Lee(1984)]{DraineLee84} Draine, B.~T.~\& Lee,
H.~M.\ 1984, \apj, 285, 89

\bibitem[Ferguson et~al.(1997)]{Ferg97}  Ferguson, J.~W., Korista, K.~T.,
Baldwin, J.~A., \& Ferland, G.~J.\ 1997, \apj, 487, 122

\bibitem[Grevesse \& Sauval(1998)]{Grevesse98} Grevesse, N.~\&
Sauval, A.~J.\ 1998, Space Science Reviews, 85, 161

\bibitem[Groves, Dopita \& Sutherland (2003)]{Groves03} Groves, B. A.,
Dopita, M. A., \& Sutherland, R. S., 2003, \apjs, submitted (Paper 2)

\bibitem[Groves et al.(2004)]{map3} Groves, B. A.,
Kewley, L.~J., Dopita, M. A., Sutherland, R. S., Evans, I., Binette,
L., \& Allen, M., 2004, in preparation

\bibitem[Honda et al.(2003)]{Honda03} Honda, M., Kataza, H.,
Okamoto, Y.~K., Miyata, T., Yamashita, T., Sako, S., Takubo, S., \& Onaka,
T.\ 2003, \apjl, 585, L59

\bibitem[Hony, Bouwman, Keller, \& Waters(2002)]{Hony02} Hony,
S., Bouwman, J., Keller, L.~P., \& Waters, L.~B.~F.~M.\ 2002, \aap, 393,
L103

\bibitem[Hony, Tielens, Waters, \& de Koter(2003)]{Hony03}
Hony, S., Tielens, A.~G.~G.~M., Waters, L.~B.~F.~M., \& de Koter, A.\
2003,
\aap, 402, 211

\bibitem[J{\" a}ger et al.(2003)]{Jager03} J{\" a}ger, C.,
Fabian, D., Schrempel, F., Dorschner, J., Henning, T., \& Wesch, W.\ 2003,
\aap, 401, 57

\bibitem[Jenkins(1987)]{Jenkins87} Jenkins, E.~B.\ 1987, ASSL
Vol.~134: Interstellar Processes, 533

\bibitem[Jones, Tielens \& Hollenbach(1996)]{Jones96} Jones,A.~P.,
Tielens, A.~G.~G.~M. \& Hollenbach, D.~J. 1996, \apj, 469, 740

\bibitem[Kaiser et al.(2000)]{Kaiser00} Kaiser, M.~E.~et al.\ 
2000, \apj, 528, 260

\bibitem[Kennicutt, Bresolin, \& Garnett(2003)]{Kennicutt03} 
Kennicutt, R.~C., Bresolin, F., \& Garnett, D.~R.\ 2003, \apj, 591,
801

\bibitem[Komossa \& Schulz(1997)]{KomSch97}  Komossa, S.~\& Schulz, H.\
1997, \aap, 323, 31

\bibitem[Koski(1978)]{Koski78}
Koski, A.~T. 1978, \apj, 223, 56

\bibitem[Kraemer \& Crenshaw(2000)]{KC2000}  Kraemer, S. B., \& Crenshaw,
D.
M., 2000, \apj, 532, 266

\bibitem[Kraemer et al.(2000)]{Kraemer00} Kraemer, S.~B., 
Crenshaw, D.~M., Hutchings, J.~B., Gull, T.~R., Kaiser, M.~E., Nelson, 
C.~H., \& Weistrop, D.\ 2000, \apj, 531, 278

\bibitem[Kraemer et~al.(1994)]{Kraemer94} 
Kraemer, S.~B., Wu, C., Crenshaw, D.~M., \& Harrington, J.~P.\ 1994, \apj, 
435, 171

\bibitem[Laor \& Draine(1993)]{Laor93} Laor, A.~\& Draine,
B.~T.\ 1993, \apj, 402, 441

\bibitem[Laor et al.(1997)]{Laor97} Laor, A., Fiore, F., 
Elvis, M., Wilkes, B.~J., \& McDowell, J.~C.\ 1997, \apj, 477, 93

\bibitem[Mathis, Rumpl, \& Nordsieck(1977)]{MRN77} Mathis,
J.~S., Rumpl, W., \& Nordsieck, K.~H.\ 1977, \apj, 217, 425 (MRN)

\bibitem[Maciel \& Pottasch(1982)]{Maciel82} Maciel, W.~J.~\&
Pottasch, S.~R.\ 1982, \aap, 106, 1

\bibitem[Mouhcine \& Contini(2002)]{Mouhcine02} Mouhcine, M.~\&
Contini, T.\ 2002, \aap, 389, 106

\bibitem[Moorwood et al.(1996)]{moor96}  Moorwood, A. F. M., Lutz, D.,
Oliva, E., Marconi, A., Netzer, H., Genzel, R., Sturm, E., \& de Graauw,
T.,
1996, \aap, 315, L109

\bibitem[Netzer \& Laor(1993)]{Netzer93} Netzer, H.~\& Laor, A.\ 
1993, \apjl, 404, L51

\bibitem[Ogle  et al.(2003)]{Ogle03} Ogle,P.~M., Brookings,T.,
Canizares,C.~R.; Lee,J.~C. \& Marshall,H.L. 2003, \aa, 402, 849

\bibitem[Oliva et al.(1994)]{oliv94}  Oliva, E., Salvati, M., Moorwood, A.
F. M., \& Marconi, A., 1994, \aap, 288, 457

\bibitem[Onaka \& Okada(2003)]{Onaka03} Onaka, T.~\& Okada, Y.\
2003, \apj, 585, 872

\bibitem[Osterbrock(1989)]{Ost89}  Osterbrock, D. E., 1989, Astrophysics
of
Gaseous Nebulae and Active Galactic Nuclei, (University Science Books)

\bibitem[Osterbrock, Tran \& Veilleux(1992)]{Osterbrock92}
Osterbrock, D.E., Tran, H.D., \& Veilleux, S. (1992), \apj, 389, 196

\bibitem[Pagel, Simonson, Terlevich, \& Edmunds(1992)]{Pagel92}
Pagel, B.~E.~J., Simonson, E.~A., Terlevich, R.~J., \& Edmunds, M.~G.\
1992, \mnras, 255, 325 

\bibitem[Pedlar et al.(1989)]{Pedlar89} Pedlar, A., Meaburn, J., 
Axon, D.~J., Unger, S.~W., Whittle, D.~M., Meurs, E.~J.~A., Guerrine, N., 
\& Ward, M.~J.\ 1989, \mnras, 238, 863

\bibitem[Pitman, Clayton, \& Gordon(2000)]{Pitman00} Pitman, 
K.~M., Clayton, G.~C., \& Gordon, K.~D.\ 2000, \pasp, 112, 537 

\bibitem[Prieto \& Viegas(2000)]{Prieto00} Prieto, M.~A.~\& 
Viegas, S.~M.\ 2000, \apj, 532, 238

\bibitem[Radomski et al.(2003)]{Radomski03} Radomski, J.~T., Pi{\~ 
n}a, R.~K., Packham, C., Telesco, C.~M., De Buizer, J.~M., Fisher, R.~S., 
\& Robinson, A.\ 2003, \apj, 587, 117

\bibitem[Russell \& Bessell(1989)]{Russell89} Russell, S.~C.~\& 
Bessell, M.~S.\ 1989, \apjs, 70, 865 

\bibitem[Russell \& Dopita(1990)]{Russell90} Russell, S.~C.~\& 
Dopita, M.~A.\ 1990, \apjs, 74, 93 

\bibitem[Russell \& Dopita(1992)]{Russell92} Russell, S.~C.~\& 
Dopita, M.~A.\ 1992, \apj, 384, 508 

\bibitem[Savage \& Sembach(1996)]{SavageS96} Savage, B.~D.~\&
Sembach, K.~R.\ 1996, \araa, 34, 279

\bibitem[Sazonov, Ostriker, \& Sunyaev(2004)]{Sazanov04} Sazonov, S.~Y., Ostriker, J.~P., \& Sunyaev, R.~A.\ 2004, \mnras, 347, 144

\bibitem[Stasinska(1984)]{Stasinska84}
Stasinska, G. (1984), \aa, 13, 341

\bibitem[Sutherland \& Dopita(1993)]{Sutherland93}Sutherland, R.~S., \&
Dopita, M.~A. 1993, \apjs, 88, 253

\bibitem[Tomono, Doi, Usuda, \& Nishimura(2001)]{Tomono01} 
Tomono, D., Doi, Y., Usuda, T., \& Nishimura, T.\ 2001, \apj, 557, 637 

\bibitem[Van Kerckhoven, Tielens, \& Waelkens(2002)]{VanK02}
Van Kerckhoven, C., Tielens, A.~G.~G.~M., \& Waelkens, C.\ 2002, \aap,
384, 568

\bibitem[Veilleux \& Osterbrock(1987)]{Veilleux87}
Veilleux, S., \& Osterbrock, D.E. (1987), \apjs, 63, 295

\bibitem[V\'eron-Cetty \& V\'eron(2000)]{VV00}  V\'{e}ron-Cetty, M. P. \&
V\'{e}ron, P., 2000, \aapr, 10, 81

\bibitem[Weingartner \& Draine(2001a)]{Weingartner01a} Weingartner,
J.~C.~\& Draine, B.~T.\ 2001a, \apj, 548, 296

\bibitem[Weingartner \& Draine(2001b)]{Weingartner01b} Weingartner,
J.~C.~\& Draine, B.~T.\ 2001b, \apjs, 134, 263

\bibitem[Whittle(1996)]{Whittle96} Whittle, M. 1996, \apjs, 79, 49


\end{thebibliography}
\end{document}